\documentclass[a4paper]{article}%
\usepackage[top=50pt,bottom=50pt,left=60pt,right=60pt]{geometry}

\usepackage{amsfonts}
\usepackage{amsmath}
\usepackage{mathtools} 
\mathtoolsset{showonlyrefs=true}
\usepackage[ntheorem, framemethod=TikZ]{mdframed} 
\usepackage{framed}  
\usepackage[amsmath, hyperref, framed]{ntheorem}%
\usepackage{mathdots}
\usepackage{amssymb}

\usepackage[labelfont=bf,textfont=normalfont,labelformat=simple,labelsep=colon,aboveskip=2pt,belowskip=3pt]{caption}
\usepackage[labelfont=bf,textfont=normalfont,labelformat=simple,labelsep=colon,aboveskip=0pt,belowskip=10pt]{subcaption}
\usepackage{psfrag,graphicx}
\usepackage{epsfig}
\usepackage{epstopdf}
\usepackage{float}%
\usepackage{color}

\usepackage[textwidth=1.2cm, textsize=scriptsize]{todonotes}

\usepackage{hyperref}%
\hypersetup{%
  breaklinks=true,%
  linktocpage=true,
  colorlinks=false,%
  linkcolor=blue,%
  citecolor=blue,%
  filecolor=blue,%
  urlcolor=blue,%
  pdftitle={},%
  pdfsubject={},%
  pdfauthor={},%
  pdfkeywords={},%
  pdfproducer={LaTeX with hyperref}%
  }%

\usepackage{ifthen}%
\usepackage{mfirstuc}
\usepackage{xkeyval}%
\usepackage{xfor}%
\usepackage{amsgen}%
\usepackage{etoolbox}%
\usepackage{datatool-base}%

\usepackage[%
  xindy,%
  style=long,%
  toc=true,%
  nonumberlist,%
  acronym,%
  acronymlists={ensemble_abbreviations},%
  hyperfirst=true
  ]{glossaries}%

\newacronym{PMCMC}{PMCMC}{particle Markov chain Monte Carlo}%
\newacronym{EMCMC}{EMCMC}{ensemble Markov chain Monte Carlo}%
\newacronym{APF}{APF}{auxiliary particle filter}%
\newacronym{FAAPF}{FA-APF}{fully-adapted auxiliary particle filter}%
\newacronym{PF}{PF}{particle filter}%
\newacronym{EHMM}{EHMM}{embedded hidden Markov models}%
\newacronym{HMM}{HMM}{hidden Markov model}%
\newacronym{CPF}{CPF}{conditional particle filter}%
\newacronym{CPFBS}{CPF-BS}{conditional particle filter with backward sampling}%
\newacronym{CPFAS}{CPF-AS}{conditional particle filter with backward sampling}%
\newacronym{PG}{PG}{particle Gibbs}%
\newacronym{PGBS}{PG-BS}{particle Gibbs sampler with backward sampling}%
\newacronym{PGAS}{PG-AS}{particle Gibbs sampler with backward sampling}%
\newacronym{SQMC}{SQMC}{sequential quasi Monte Carlo}%
\newacronym{RQMC}{RQMC}{randomised quasi Monte Carlo}%
\newacronym[user1={ancestor-sampling}]{AS}{AS}{ancestor sampling}%
\newacronym[user1={backward-sampling}]{BS}{BS}{backward sampling}%
\newacronym{PDF}{PDF}{probability density function}%
\newacronym{IID}{IID}{independent and identically distributed}%
\newacronym{MCMC}{MCMC}{Markov chain Monte Carlo}%
\newacronym{MH}{MH}{Metropolis--Hastings}%
\newacronym{ESS}{ESS}{effective sample size}%
\newacronym{PMMH}{PMMH}{particle marginal Metro\-po\-lis--Has\-tings}%
\newacronym{MCWM}{MCWM}{Monte Carlo within Metropolis}%
\newacronym{CDF}{CDF}{cumulative distribution function}%
\newacronym{SMC}{SMC}{sequential Monte Carlo}%
\newacronym{CSMC}{CSMC}{conditional sequential Monte Carlo}%
\newacronym{MCMCPF}{MCMC PF}{(bootstrap) particle filter with MCMC moves}%
\newacronym{MCMCFAAPF}{MCMC FA-APF}{fully-adapted auxiliary particle filter with MCMC moves}%
\newacronym{MCMCAPF}{MCMC APF}{auxiliary particle filter with MCMC moves}%
\newacronym{EPSRC}{EPSRC}{Engineering and Physical Sciences Research Council}%
\newacronym{ESJD}{ESJD}{expected squared jumping distance}

\usepackage{lineno}

\newcommand*\patchAmsMathEnvironmentForLineno[1]{%
  \expandafter\let\csname old#1\expandafter\endcsname\csname #1\endcsname
  \expandafter\let\csname oldend#1\expandafter\endcsname\csname end#1\endcsname
  \renewenvironment{#1}%
     {\linenomath\csname old#1\endcsname}%
     {\csname oldend#1\endcsname\endlinenomath}}%
\newcommand*\patchBothAmsMathEnvironmentsForLineno[1]{%
  \patchAmsMathEnvironmentForLineno{#1}%
  \patchAmsMathEnvironmentForLineno{#1*}}%
\AtBeginDocument{%
\patchBothAmsMathEnvironmentsForLineno{equation}%
\patchBothAmsMathEnvironmentsForLineno{align}%
\patchBothAmsMathEnvironmentsForLineno{flalign}%
\patchBothAmsMathEnvironmentsForLineno{alignat}%
\patchBothAmsMathEnvironmentsForLineno{gather}%
\patchBothAmsMathEnvironmentsForLineno{multline}%
}

\usepackage[inline]{enumitem}

\qedsymbol{\ensuremath{\Box}}
\theoremstyle{plain}%

\theoremheaderfont{\rmfamily\upshape\bfseries} 
\theorembodyfont{\normalfont\slshape}%
\theoremseparator{.}

\newtheorem{assumption}{Assumption}%
\newtheorem{remark}{Remark}%
\newtheorem{example}{Example}%

\theorembodyfont{\normalfont}%

\newmdtheoremenv[
 ntheorem=true,
 skipbelow = .6\baselineskip plus 1ex minus 1ex,
 skipabove = .6\baselineskip plus 1ex minus 1ex,
 innerleftmargin = 0pt,
 innerrightmargin = 0pt,
 leftline = false,
 rightline = false,
 needspace = 5ex 
]{framedAlgorithm}{Algorithm}

\theoremstyle{nonumberplain}%
\theoremheaderfont{\indent\rmfamily\itshape}%
\theorembodyfont{\upshape \mdseries}%
\theoremsymbol{\ensuremath{_\Box}}
\theoremsymbol{\ensuremath{_\square}}
\usepackage{csquotes}%
\usepackage{natbib}
\setcitestyle{authoryear}
\begin{document}

\title{On embedded hidden Markov models and\\particle Markov chain Monte Carlo methods}
\author{Axel Finke\\Department of Statistical Science, University College London, UK.\\Arnaud Doucet\\Department of Statistics, Oxford University, UK.\\Adam M.~Johansen\\Department of Statistics, University of Warwick, UK.}
\maketitle

\begin{abstract}
  \noindent{}The \gls{EHMM} sampling method is a \gls{MCMC} technique for state inference in non-linear non-Gaussian state-space models which was proposed in \citet{Neal2003,Neal2004} and extended in \citet{ShestopaloffNeal2016}. An extension to Bayesian parameter inference was presented in \citet{ShestopaloffNeal2013}. An alternative class of \gls{MCMC} schemes addressing similar inference problems is provided by \gls{PMCMC} methods \citep{AndrieuDoucetHolenstein2009,AndrieuDoucetHolenstein2010}. All these methods rely on the introduction of artificial extended target distributions for multiple state sequences which, by construction, are such that one randomly indexed sequence is distributed according to the posterior of interest. By adapting the \glsdesc{MH} algorithms developed in the framework of \gls{PMCMC} methods to the \gls{EHMM} framework, we obtain novel \gls{PF}-type algorithms for state inference and novel \gls{MCMC} schemes for parameter and state inference. In addition, we show that most of these algorithms can be viewed as particular cases of a general \gls{PF} and \gls{PMCMC} framework. We compare the empirical performance of the various algorithms on low- to high-dimensional state-space models. We demonstrate that a properly tuned conditional \gls{PF} with `local' \gls{MCMC} moves proposed in \citet{ShestopaloffNeal2016} can outperform the standard conditional \gls{PF} significantly when applied to high-dimensional state-space models  while the novel \gls{PF}-type algorithm could prove to be an interesting alternative to standard \glspl{PF} for likelihood estimation in some lower-dimensional scenarios.
\end{abstract}

\glsresetall
\glsunset{MCMC}

\section{Introduction}

Throughout this work, for concreteness, we will describe both \gls{PMCMC} and \gls{EHMM} methods in the context of performing inference in non-linear state-space models. However, we stress that those methods can be used to perform inference in other contexts.

Non-linear non-Gaussian state-space models constitute a popular class of time series  models which can be described in the time-homogeneous case as follows --- throughout this paper we consider the time-homogeneous case, noting that the generalisation to time-inhomogeneous models is straightforward but notationally cumbersome. Let  $\{x_t\}_{t \geq 1}$ be an $\mathcal{X}$-valued
latent Markov process satisfying
\begin{equation}
  x_1 \sim \mu_{\theta}(\,\cdot\,) \quad \text{and} \quad x_{t}\vert (x_{t-1} = x)  \sim f_{\theta}(\,\cdot\,| x), \text{ for $t \geq 2$}. \label{eq:modtransition}%
\end{equation}
and let $\{y_t\}_{t \geq 1}$ be a sequence of $\mathcal{Y}$-valued observations 
which are conditionally independent given $\{x_t\}_{t \geq 1}$
and which satisfy
\begin{equation}
  y_t \vert (x_1,\dotsc, x_t = x, x_{t+1}, \dotsc )
  \sim g_{\theta}(\,\cdot\,| x), \text{ for
  $t \geq 1$}. \label{eq:modobs}%
\end{equation}
Here $\theta\in\varTheta$ denotes the vector of parameters of the model.

Let $z_{i:j}$ denotes the components $(z_i,z_{i+1},\dotsc,z_j)
$ of a generic sequence $\{z_t\}_{t \geq 1} $.\ Assume that we have access to
a realization of the observations $Y_{1:T}=y_{1:T}$. If $\theta$ is known,
inference about the latent states $x_{1:T}$ relies upon
\begin{equation}
  p_{\theta}(x_{1:T} \vert y_{1:T}) 
  = \frac{p_{\theta}(x_{1:T}, y_{1:T})}{p_{\theta}(y_{1:T})},
\end{equation}
where
\begin{equation}
  p_{\theta} (x_{1:T}, y_{1:T}) 
  = \mu_{\theta}(x_1)
  \prod_{t=2}^{T} f_{\theta}(x_{t}\vert x_{t-1})
  \prod_{t=1}^{T} g_{\theta}(y_t \vert x_t)  .
\end{equation}
When $\theta$ is unknown, to conduct Bayesian inference a prior density $p(\theta)$ is assigned to the parameters and inference proceeds via
the joint posterior density
\begin{equation}
  p(x_{1:T},\theta \vert y_{1:T})  
  = p(\theta \vert y_{1:T})  p_{\theta}(x_{1:T}\vert y_{1:T}),
\end{equation}
where the marginal posterior distribution of the parameter satisfies
\begin{equation}
  p(\theta \vert y_{1:T})  \propto p(\theta)  p_{\theta}(y_{1:T}),
\end{equation}
the likelihood $ p_{\theta}(y_{1:T})$ being given by
\begin{equation}
  p_{\theta}(y_{1:T}) =\int  p_{\theta} (x_{1:T}, y_{1:T}) \mathrm{d}x_{1:T}.
\end{equation}

Many algorithms have been proposed over the past twenty-five years to perform inference for this class of models; see \citet{Kantas2015} for a recent survey. We focus here on the \gls{EHMM} algorithm introduced in \citet{Neal2003,Neal2004} and on \gls{PMCMC} introduced in
\cite{AndrieuDoucetHolenstein2009,AndrieuDoucetHolenstein2010}. Both classes of methods are fairly generic and do not require the state-space model under consideration to possess additional structural properties beyond \eqref{eq:modtransition} and \eqref{eq:modobs}. The \gls{EHMM} method has been recently extended in \citet{ShestopaloffNeal2013,ShestopaloffNeal2016} while extensions of \gls{PMCMC} have also been proposed in, among other works, \citet{Whiteley2010} and \citet{LindstenJordanSchon2014}. In particular, \citet{Whiteley2010} combined the conditional \gls{PF} algorithm of \cite{AndrieuDoucetHolenstein2009,AndrieuDoucetHolenstein2010} with a backward sampling step. We will denote the resulting algorithm as the conditional \gls{PF} with \gls{BS}.

\begin{figure}
 \centering
 \includegraphics{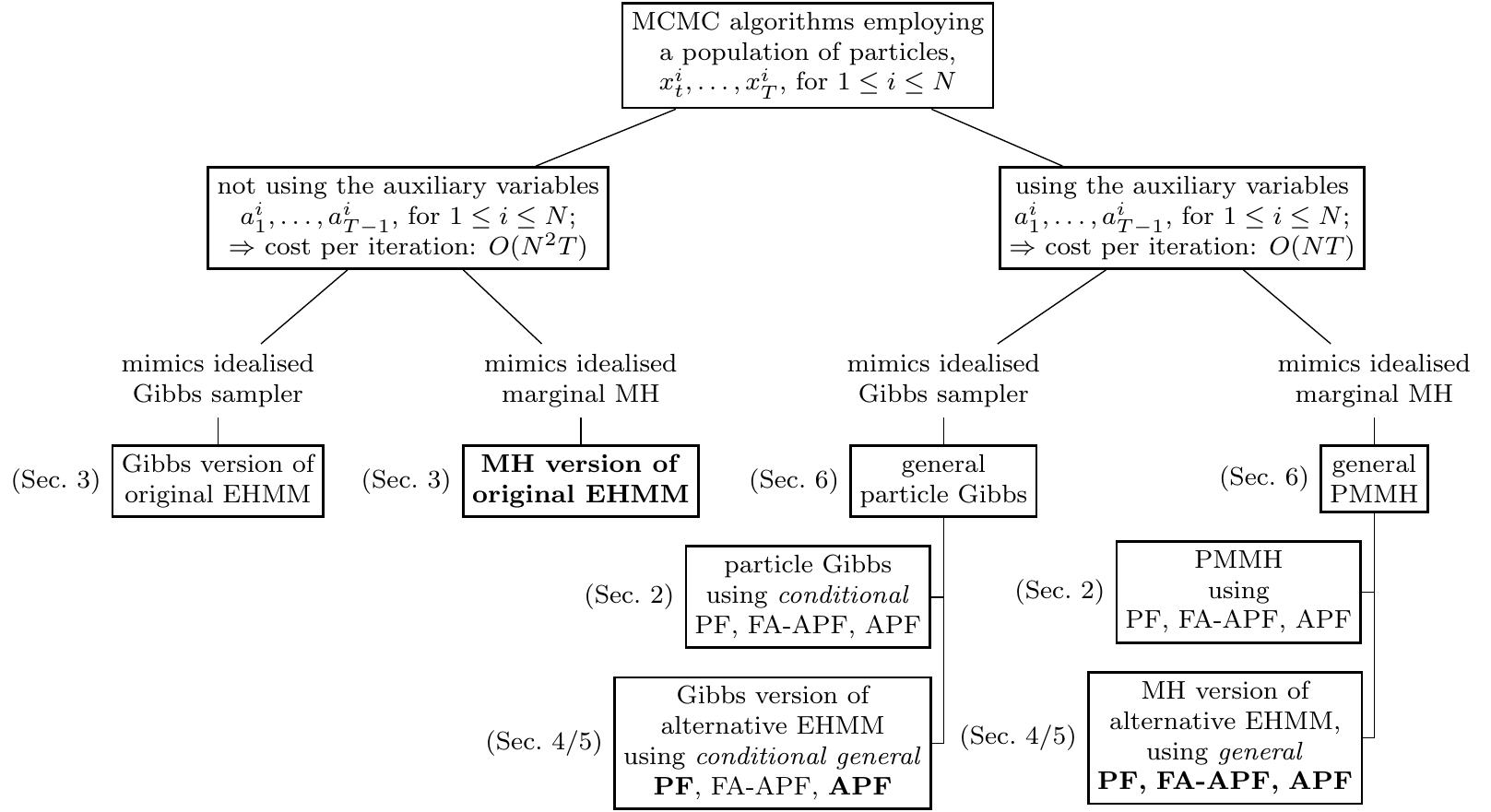}\\[2em]
 \caption{Relationship between the various classes of algorithms discussed in this work. A general construction admitting all of these as special cases can be found in \citet[Section~1.4]{Finke2015}. Novel methodology introduced in this work is highlighted in bold.}
 \label{fig:relationship_between_algorithms}
\end{figure}

Both \gls{EHMM} and \gls{PMCMC} methods rely upon sampling a population of $N$ particles for the state $x_{t}$ and introducing an extended target distribution over the resulting $N^{T}$ potential sequences $x_{1:T}$ such that one of the sequences selected uniformly at random is at equilibrium by construction. It was observed in \citet[p.~116]{LindstenSchon2013} that conditional \gls{PF} with \gls{BS} is reminiscent of the \gls{EHMM} method proposed in \citet{Neal2003,Neal2004} and some connections were made between some simple \gls{EHMM} methods and \gls{PMCMC} methods in \citet[pp.~82--87]{Finke2015} who also showed that both methods can be viewed as special cases of a much more general construction. However, to the best of our knowledge, the connections between the two classes of methods have never been investigated thoroughly. Indeed, such an analysis was deemed of interest in \citet{ShestopaloffNeal2014}, where we note that \gls{EHMM} methods are sometimes alternatively referred to as \emph{ensemble} \gls{MCMC} methods:
\begin{quote}
 ``It would \ldots{} be interesting to compare the performance of the ensemble \gls{MCMC} method with the [\gls{PMCMC}]-based methods of \cite{AndrieuDoucetHolenstein2010} and also to see whether techniques used to improve [particle \gls{MCMC}] methods can be used to improve ensemble methods and vice versa.''
\end{quote}
In this work, we characterize this relationship and show that it is possible to exploit the similarities between these methods to derive new inference algorithms. The relationship between the various classes of algorithms discussed in this work is shown in Figure~\ref{fig:relationship_between_algorithms}. The remainder of the paper is organized as follows. 

Section \ref{sec:pmcmc} reviews some \gls{PMCMC} schemes, including the \gls{PMMH} algorithm and \gls{PG} samplers. We recall how the validity of these algorithms can be established by showing that they are standard \gls{MCMC} algorithms sampling from an extended target distribution. In particular, the \gls{PMMH} algorithm can be thought of as a standard \gls{MH} algorithm sampling from this extended target using a \gls{PF} proposal for the states. Likewise, the theoretical validity of the conditional \gls{PF} with \gls{BS} can be established by showing that it corresponds to a (``partially collapsed'' -- see \citet{VanDyk2008}) Gibbs sampler \citep{Whiteley2010}. 

Section~\ref{sec:embeddedHMM} is devoted to the `original' \gls{EHMM} from \citet{Neal2003,Neal2004}. At the core of this methodology is an extended target distribution which shares common features with the \gls{PMCMC} target. We show that the \gls{EHMM} method can be reinterpreted as a collapsed Gibbs sampling procedure for this target. This provides an alternative proof of validity of this algorithm. More interestingly, it is possible to come up with an original \gls{MH} scheme to sample from this extended target distribution reminiscent of \gls{PMMH}. However, whereas the \gls{PMMH} algorithm relies on  \gls{PF} estimates of the likelihood $p_{\theta}(y_{1:T})$, this \gls{MH} version of \gls{EHMM} relies on an estimate of $p_{\theta}(y_{1:T})$ computed using a finite-state \gls{HMM}, the cardinality of the state-space being $N$. The computational cost of both of these original \gls{EHMM} methods is $O(N^2T)$ in contrast to the $O(NT)$-cost of \gls{PMCMC} methods.

The high computational cost of the original \gls{EHMM} method has partially motivated the development of a novel class of alternative \gls{EHMM} methods which bring the computational complexity down to $O(NT)$. As described in Section~\ref{sec:embeddedHMMnewversion}, this is done by introducing a set of auxiliary variables playing the same r\^ole as the ancestor indices generated in the resampling step of a standard \gls{PF}. This leads to the extended target distribution introduced in \citet{ShestopaloffNeal2016}. We show that this target coincides in a special case with the extended target of \gls{PMCMC} when one uses the \gls{FAAPF} \citep{PittShephard1999} and the resulting \gls{EHMM} coincides with the conditional \gls{FAAPF} with \gls{BS} in this scenario. We show once more that the validity of this novel \gls{EHMM} method can be established by using a collapsed Gibbs sampler. 

In Section~\ref{sec:novel_methodology}, we derive several novel, practical extensions to the alternative \gls{EHMM} method. First, we show that the alternative \gls{EHMM} framework can also be used to derive an \gls{MH} algorithm which, once again, is very similar to the \gls{PMMH} algorithm except that $p_{\theta}(y_{1:T})$ is estimated unbiasedly using a novel  \gls{PF} type algorithm relying on local \gls{MCMC} moves. Second, we derive additional bootstrap \gls{PF} and general \gls{APF} type variants of the alternative \gls{EHMM} method.  

In Section~\ref{sec:general_pmcmc}, we describe a general, unifying \gls{PMCMC} framework which admits all variants of standard \gls{PMCMC} methods and all variants of alternative \gls{EHMM} discussed in this work as special cases. This also allows us to generalize the \glsdesc{AS} scheme from \citet{LindstenJordanSchon2014}.

In Section~\ref{sec:simulations}, we empirically compare the performance of all the algorithms mentioned above. Our results indicate that, as suggested in \citet{ShestopaloffNeal2016}, a properly tuned version of the conditional \gls{PF} (and hence \gls{PG} sampler) using \gls{MCMC} moves proposed in \citet{ShestopaloffNeal2016} can outperform existing methods in high dimensions while the (`non-conditional') \glspl{PF} using \gls{MCMC} moves are a potentially interesting alternative to standard \glspl{PF} for likelihood and state estimation for lower-dimensional models.

\section{Particle Markov chain Monte Carlo methods}
\label{sec:pmcmc}

This section reviews \gls{PMCMC} methods. For transparency, we first restrict ourselves in this section to the scenario in which the underlying \gls{PF} used is the bootstrap \gls{PF}, and then discuss the \glsdesc{FAAPF} before finally considering the case of general \glsdescplural{APF}.

\subsection{Extended target distribution}

Let $N$ be an integer such that $N \geq 2$. \gls{PMCMC} methods rely on the following extended target density on $\varTheta \times \mathcal{X}^{NT}\times \{1,\dotsc,N\}^{N{(T-1)}+1}$
\begin{equation}
  \tilde{\pi}{\bigl(\theta,b_{1:T},\mathbf{x}_{1:T},\mathbf{a}_{1:T-1}^{-b_{2:T}}\bigr)}  
  \coloneqq
   \frac{1}{N^T} \times \underbrace{\pi{\bigl(\theta,x_{1:T}^{b_{1:T}}\bigr)}}_{\mathclap{\text{\footnotesize{target}}}} \times \underbrace{\phi_\theta{\bigl(\mathbf{x}_{1:T}^{-b_{1:T}},\mathbf{a}_{1:T-1}^{-b_{2:T}}|x_{1:T}^{b_{1:T}},b_{1:T}\bigr)}  }_{\mathclap{\text{\footnotesize{law of conditional \gls{PF}}}}}, \label{eq:PMCMCtarget}%
\end{equation}
where $\smash{\pi(\theta,x_{1:T})  \coloneqq p(x_{1:T},\theta| y_{1:T})}$ represents the posterior distribution of interest. In addition, the \emph{particles} $\mathbf{x}_t \coloneqq \{  x_t^1,\dotsc,x_t^N \}  \in\mathcal{X}^N$, \emph{ancestor indices} $\mathbf{a}_t \coloneqq \{a_t^1,\dotsc,a_t^N\} \in \{1,\dotsc,N\}^N$ and \emph{particle indices} $b_{1:T} \coloneqq \{  b_1,\dotsc,b_T\}$ are related as
\begin{equation}
 \mathbf{x}_t^{-b_t}=\mathbf{x}_t\backslash x_t^{b_t}, \quad \mathbf{x}_{1:T}^{-b_{1:T}}=\bigl\{  \mathbf{x}_1^{-b_1},\dotsc,\mathbf{x}_T^{-b_T}\bigr\}, \quad \mathbf{a}_{t-1}^{-b_t}=\mathbf{a}_{t-1}\backslash a_{t-1}^{b_t}, \quad \mathbf{a}_{1:T-1}^{-b_{2:T}}=\bigl\{  \mathbf{a}_1^{-b_{2}},\dotsc,\mathbf{a}_{T-1}^{-b_T}\bigr\}.
\end{equation}
In particular, given $b_T$, the particle indices $b_{1:T-1}$ are deterministically related to the ancestor indices by the recursive relationship
\begin{equation}
 b_t = a_t^{b_{t+1}}, \quad \text{for $t = T-1, \dotsc, 1$.}
\end{equation} 
Finally, for any $\smash{(x_{1:T}^{b_{1:T}},b_{1:T}) \in\mathcal{X}^N \times \bigl\{1,\dotsc,N\}^N}$, $\phi_\theta$ denotes a conditional distribution induced by an algorithm referred to as a \gls{CPF}
\begin{equation}
  \phi_\theta{\bigl(\mathbf{x}_{1:T}^{-b_{1:T}}, \mathbf{a}_{1:T-1}^{-b_{2:T}}\big| x_{1:T}^{b_{1:T}},b_{1:T}\bigr)} 
  \coloneqq 
  \prod_{\substack{\mathllap{i}=\mathrlap{1}\\\mathllap{i} \neq \mathrlap{b_1}}}^N
  \mu_\theta{\bigl(  x_1^i\bigr)} \prod_{t=2}^T \prod_{\substack{\mathllap{i}=\mathrlap{1}\\\mathllap{i} \neq \mathrlap{b_t}}}^N w_{\theta,t-1}^{a_{t-1}^i}\,f_\theta{\bigl( x_t^i\big|x_{t-1}^{a_{t-1}^i}\bigr)}, \label{eq:CPF}
\end{equation}
where
\begin{equation}
  w_{\theta,t}^i
  \coloneqq \frac{g_\theta(y_t |x_t^i)}{\sum_{j=1}^N g_\theta(y_t | x_t^j)} \label{eq:multinomialresampling}%
\end{equation}
represents the normalised weight associated with the $i$th particle at time~$t$.

The key feature of this high-dimensional target is that by construction it ensures that $\smash{(  \theta,x_{1:T}^{b_{1:T}})}$ is distributed according to the posterior of interest. \gls{PMCMC} methods are \gls{MCMC} algorithms which sample from this extended target, hence from the posterior of interest.

\subsection{Particle marginal Metropolis--Hastings}
\label{subsec:pmmh}

\glsreset{PMMH}
\glsreset{MH}

The \emph{\gls{PMMH}} algorithm is a \gls{MH} algorithm targeting $\tilde{\pi}(
\theta,b_{1:T},\mathbf{x}_{1:T},\mathbf{a}_{1:T-1}^{-b_{2:T}})  $
defined through \eqref{eq:PMCMCtarget}, \eqref{eq:CPF} and \eqref{eq:multinomialresampling} using a proposal of the form
\begin{equation}
  q{(\theta,{\theta'})} \times 
  \underbrace{\Psi_{{\theta'}}(  \mathbf{x}_{1:T},\mathbf{a}_{1:T-1})}_{\mathclap{\text{\footnotesize{law of \gls{PF}}}}} \times \underbrace{w_{{\theta'},T}^{b_T}}_{\mathclap{\text{\parbox{1.3cm}{\centering\footnotesize{path selection}}}}}, \label{eq:proposalparticlefilter}%
\end{equation}
where $b_{1:T}$ is again obtained via the reparametrisation $\smash{b_t = a_t^{b_{t+1}}}$ for $t = T-1, \dotsc,1$ and $\Psi_\theta(\mathbf{x}_{1:T},\mathbf{a}_{1:T-1})$ is the law induced by a bootstrap \gls{PF}
\begin{equation}
  \Psi_\theta(  \mathbf{x}_{1:T},\mathbf{a}_{1:T-1})
  \coloneqq
  \prod_{i=1}^N \mu_\theta(  x_1^i)
  \prod_{t=2}^T \prod_{i=1}^N w_{\theta,t-1}^{a_{t-1}^i} f_\theta\bigl(x_t^i \big| x_{t-1}^{a_{t-1}^i}\bigr). \label{eq:distributionPF}%
\end{equation}
The resulting \gls{MH} acceptance probability is of the form
\begin{equation}
  1 \wedge 
  \frac{\hat{p}_{{\theta'}}(y_{1:T})p({\theta'})}{\hat{p}_\theta(y_{1:T})p(\theta)}
  \frac{q({\theta'},\theta)}{q(\theta,{\theta'})}, \label{eq:PMMHratio}%
\end{equation}
where
\begin{equation}
  \hat{p}_\theta(y_{1:T}) 
  \coloneqq \prod_{t=1}^T \biggl[\frac{1}{N} \sum_{i=1}^N g_\theta(y_t|
  x_t^i) \biggr]\label{eq:likelihoodestimatorPF}
\end{equation}
is well known to be an unbiased estimate of $p_\theta(y_{1:T})$; see \cite{DelMoral2004}. We stress that the unbiased estimates appearing in the numerator and denominator of \eqref{eq:PMMHratio} each depends upon the particles (and ancestor indices) generated in distinct \glspl{PF} but we suppress this dependence to keep the notation as simple as is possible. The validity of the expression in \eqref{eq:PMMHratio} follows directly by noting that:
\begin{align}
  \frac{\tilde{\pi}(  \theta,b_{1:T},\mathbf{x}_{1:T},\mathbf{a}_{1:T-1}^{-b_{2:T}}) }{\Psi_\theta(  \mathbf{x}_{1:T},\mathbf{a}_{1:T-1})  w_{\theta,T}^{b_T}}  
  & = \frac{1}{N^T} \frac{\pi(  \theta,x_{1:T}^{b_{1:T}})}{\mu_\theta(x_1^{b_1}) \bigl[\prod_{t=2}^T w_{\theta,t-1}^{b_{t-1}} f_\theta(x_t^{b_t} | x_{t-1}^{b_{t-1}}) \bigr] w_{\theta,T}^{b_T}}\\
  & = \frac{p(  \theta| y_{1:T})  }{N^T}\frac
  {\mu_\theta(  x_1^{b_1}) g_\theta(  y_1| x_1^{b_1}) \prod_{t=2}^T f_\theta(  x_t^{b_t}| x_{t-1}^{b_{t-1}})
  g_\theta(  y_t| x_t^{b_t})  }{\mu_\theta(  x_1^{b_1})
  \bigl[\prod_{t=2}^T w_{\theta,t-1}^{b_{t-1}} f_\theta(x_t^{b_t} | x_{t-1}^{b_{t-1}}) \bigr] w_{\theta,T}^{b_T}}\\
  &  = p(\theta | y_{1:T}) \frac{\hat{p}_\theta(  y_{1:T})  }{p_\theta(y_{1:T})}\\
  &  \propto\hat{p}_\theta(y_{1:T})  p(\theta), \label{eq:PMCMCratiotargetproposal}%
\end{align}
where we have again used that $b_t = a_t^{b_{t+1}}$, for $t = T-1,\dotsc,1$ and that $p(\theta|y_{1:T})/p_\theta(y_{1:T}) = p(\theta)/p(y_{1:T})$; see also \cite[Theorem 2]{AndrieuDoucetHolenstein2010}.

\subsection{Particle Gibbs samplers}
\label{subsec:cpf}
\glsreset{PG}
\glsreset{BS}
\glsreset{AS}

To sample from $\pi(\theta,x_{1:T})$, one can use the \emph{\gls{PG}} sampler. The \gls{PG} sampler mimics the block Gibbs sampler iterating draws from $\pi(\theta | x_{1:T})$ and $\pi(x_{1:T}| \theta)  $. As sampling from $\pi(x_{1:T} | \theta)  $ is typically
impossible, we can use a so called conditional \gls{PF} kernel with \gls{BS} to emulate sampling from it. Given a current value of $x_{1:T}$, we perform the following steps (see \citet{AndrieuDoucetHolenstein2009}, \citet[Section~4.5]{AndrieuDoucetHolenstein2010});
\begin{enumerate}
 \item Sample $b_{1:T}$ uniformly at random and set $x_{1:T}^{b_{1:T}}\leftarrow x_{1:T}$.
 \item Run the conditional \gls{PF}, i.e.\ sample from $\phi_\theta(\mathbf{x}_{1:T}^{-b_{1:T}},\mathbf{a}_{1:T-1}^{-b_{2:T}} | x_{1:T}^{b_{1:T}},b_{1:T})$.
 \item\label{alg:simple_pg_last_step} Sample $b_T$ according to $\Pr(b_T=m) = w_\theta^{m}$ and set $b_t=a_t^{b_{t+1}}$ for $t = T-1, \dotsc, 1$.
\end{enumerate}
It was noticed in \citet{Whiteley2010} that it is possible to improve Step~\ref{alg:simple_pg_last_step}: for $t=T-1,\dotsc,1$, instead of deterministically setting $b_t=a_t^{b_{t+1}}$, one can use a backward sampling step which samples
\begin{equation}
  \Pr{\bigl(b_t=m\bigr)} \propto w_{\theta,t}^{m}f_\theta{\bigl(x_{t+1}^{b_{t+1}}\bigr| x_t^{m}\bigr)}. \label{eq:backwardsampling}
\end{equation}
To establish the validity of this procedure (i.e.\ of the conditional \gls{PF} with \gls{BS}), it was shown that this procedure is a (partially) collapsed Gibbs sampler of invariant distribution $\smash{\tilde{\pi}(b_{1:T},\mathbf{x}_{1:T},\mathbf{a}_{1:T-1} | \theta)}$, sampling recursively from $\smash{\tilde{\pi}(b_t | \theta,\mathbf{x}_{1:t},\mathbf{a}_{1:t-1},x_{t+1:T}^{b_{t+1:T}},b_{t+1:T})}$, for $t=T,T-1,\dotsc,1$. Indeed, we have
\begin{align}
 \MoveEqLeft \tilde{\pi}{\bigl(b_t\bigr| \theta,\mathbf{x}_{1:t},\mathbf{a}_{1:t-1},x_{t+1:T}^{b_{t+1:T}},b_{t+1:T}\bigr)}\\
  & \propto \sum_{b_{1:t-1}}\sum_{\mathbf{a}_{t:T-1}} \idotsint
  \frac{\pi{\bigl(  \theta,x_{1:T}^{b_{1:T}}\bigr)}  }{N^T} \prod_{\substack{\mathllap{i}=\mathrlap{1}\\\mathllap{i} \neq \mathrlap{b_1}}}^N \mu_\theta{\bigl(  x_1^i\bigr)} \smashoperator{\prod_{n=2}^T} \prod_{\substack{\mathllap{i}=\mathrlap{1}\\\mathllap{i} \neq \mathrlap{b_n}}}^N
  w_{\theta,n-1}^{a_{n-1}^i}f_\theta{\bigl(x_n ^i\bigr| x_{n-1}^{a_{n-1}^i}\bigr)}\,\mathrm{d} \mathbf{x}_{t+1:T}^{-b_{t+1:T}}\\
  & \propto \smashoperator{\sum_{b_{1:t-1}}} \pi{\bigl(  \theta,x_{1:T}^{b_{1:T}}\bigr)}
  \prod_{\substack{\mathllap{i}=\mathrlap{1}\\\mathllap{i} \neq \mathrlap{b_1}}}^N \mu_\theta{\bigl(  x_1^i\bigr)} \prod_{n=2}^t
  \prod_{\substack{\mathllap{i}=\mathrlap{1}\\\mathllap{i} \neq \mathrlap{b_n}}}^N w_{\theta,n-1}^{a_{n-1}^i}f_\theta{\bigl(x_n^i\bigr| x_{n-1}^{a_{n-1}^i}\bigr)} \\
  & = \smashoperator{\sum_{b_{1:t-1}}} \pi{\bigl(\theta,x_{1:T}^{b_{1:T}}\bigr)} \frac{\prod_{i=1}^N \mu_\theta{\bigl(  x_1^i\bigr)} \prod_{n=2}^t \prod_{i=1}^N w_{\theta,n-1}^{a_{n-1}^i}f_\theta{\bigl(x_n^i\bigr| x_{n-1}^{a_{n-1}^i}\bigr)}  }{\mu_\theta{\bigl(x_1^{b_1}\bigr)} \prod_{n=2}^t w_{\theta,n-1}^{b_{n-1}}f_\theta{\bigl(x_n^{b_n}\bigr| x_{n-1}^{b_{n-1}}\bigr)}  }, \quad \text{as $\smash{a_{n-1}^{b_n} = b_{n-1}}$,}\label{eq:CPFBS}\\
  & \propto \smashoperator{\sum_{b_{1:t-1}}} f_\theta{\bigl(x_{t+1}^{b_{t+1} }\bigr| x_t^{b_t}\bigr)}  w_{\theta,t}^{b_t}\\
  & \propto f_\theta{\bigl(x_{t+1}^{b_{t+1}}\bigr| x_t^{b_t}\bigr)}  w_{\theta,t}^{b_t},
\end{align}
where we have used that the numerator of the ratio appearing in \eqref{eq:CPFBS} is independent of $b_{1:t-1}$. 

\subsection{Extension to the fully-adapted auxiliary particle filter}
\label{Section:perfectadaptationPF}

\glsreset{FAAPF}

It is straightforward to employ a more general class of \glspl{PF} in a \gls{PMCMC} context. One such \gls{PF} is the \emph{\gls{FAAPF}}  \citep{PittShephard1999} whose incorporation within \gls{PMCMC} was explored in \citet{Pitt2012}. It is described in this subsection.  

When it is possible to sample from $p_\theta(x_1|y_1)  \propto \mu_\theta(x_1)  g_\theta(y_1| x_1)  $ and $p_\theta(x_t| x_{t-1},y_t)  \propto f_\theta(x_t| x_{t-1}) g_\theta(y_t|x_t)$ and to compute $p_\theta(y_1) = \int \mu_\theta(x_1)  g_\theta(y_1|x_1) \mathrm{d}x_1$ and $p_\theta(y_t|x_{t-1}) = \int f_\theta(x_t|x_{t-1})  g_\theta(y_t| x_t) \mathrm{d} x_t$, it is possible to\ define the target distribution $\smash{\tilde{\pi}(\theta,b_{1:T},\mathbf{x}_{1:T},\mathbf{a}_{1:T-1}^{-b_{2:T}})}$ using an alternative conditional \gls{PF} -- the conditional \gls{FAAPF} -- in \eqref{eq:PMCMCtarget} (more precisely, in these circumstances one can implement the associated \gls{PF}):
\begin{equation}
  \phi_\theta{\bigl(\mathbf{x}_{1:T}^{-b_{1:T}},\mathbf{a} _{1:T-1}^{-b_{2:T}}\bigr| x_{1:T}^{b_{1:T}},b_{1:T}\bigr)}  
  = \prod_{\substack{\mathllap{i}=\mathrlap{1}\\\mathllap{i} \neq \mathrlap{b_1}}}^N p_\theta{\bigl(x_1^i\bigr| y_1\bigr)} \prod_{t=2}^T \prod_{\substack{\mathllap{i}=\mathrlap{1}\\\mathllap{i} \neq \mathrlap{b_t}}}^N w_{\theta,t-1}^{a_{t-1}^i}p_\theta{\bigl(x_t^i\bigr| x_{t-1}^{a_{t-1}^i},y_t\bigr)},  \label{eq:CPFperfectadaptation}%
\end{equation}
where 
\begin{equation}
  w_{\theta,t}^{i} \coloneqq \frac{p_\theta(y_{t+1} | x_{t}^{i})}{\sum_{j=1}^Np_\theta(y_{t+1}|
  x_{t}^j)}. \label{eq:perfectadaptationweight}%
\end{equation}
In this case, we can target the extended distribution $\smash{\tilde{\pi}(\theta,b_{1:T},\mathbf{x}_{1:T},\mathbf{a}_{1:T-1}^{-b_{2:T}})}$ defined through \eqref{eq:PMCMCtarget}, \eqref{eq:CPFperfectadaptation} and \eqref{eq:perfectadaptationweight} using a
\gls{MH} algorithm with proposal
\begin{equation}
  q{\bigl(  \theta,{\theta'}\bigr)} \times \underbrace{\Psi_{{\theta'}}{\bigl(\mathbf{x}_{1:T},\mathbf{a}_{1:T-1}\bigr)}}_{\mathclap{\text{\footnotesize{law of \gls{FAAPF}}}}} \times \underbrace{\frac{1}{N}}_{\mathclap{\text{\parbox{1.3cm}{\centering\footnotesize{path selection}}}}},
\end{equation}
i.e.\ we pick $b_T$ uniformly at random, then set $\smash{b_t=a_t^{b_{t+1}}}$ for $t=T-1,\dotsc,1$ and $\Psi_\theta{\bigl(\mathbf{x}_{1:T},\mathbf{a}_{1:T-1}\bigr)}$ is the distribution associated with the \gls{FAAPF} instead of the bootstrap \gls{PF}
\begin{equation}
  \Psi_\theta{\bigl(\mathbf{x}_{1:T},\mathbf{a}_{1:T-1}\bigr)} 
  = \prod_{i=1}^N p_\theta{\bigl(x_1^i\bigr| y_1\bigr)} \prod_{t=2}^T \prod_{i=1}^N w_{\theta,t-1}^{a_{t-1}^i}p_\theta{\bigl(x_t^i\bigr| x_{t-1}^{a_{t-1}^i},y_t\bigr)}  .
  \label{eq:distributionperfectadaptation}%
\end{equation}
It is easy to check that the resulting \gls{MH} acceptance probability is also of the form given in \eqref{eq:PMMHratio} but with
\begin{equation}
  \hat{p}_\theta(y_{1:T}) 
   = p_\theta(y_1) \prod_{t=2}^T \biggl[\frac{1}{N} \sum_{i=1}^N p_\theta(y_t|x_{t-1}^i)\biggr]. \label{eq:likelihoodestimatorperfectadaptationPF}
\end{equation}
The conditional \gls{FAAPF} with \gls{BS} proceeds by first running the conditional \gls{FAAPF} defined in \eqref{eq:CPFperfectadaptation}, then sampling $b_T$ uniformly at random and finally sampling $b_{T-1},\dotsc,b_1$ backwards using
\begin{equation}
  \tilde{\pi}\bigl(b_t \big| \theta,\mathbf{x}_{1:t},\mathbf{a}_{1:t-1}, x_{t+1:T}^{b_{t+1:T}},b_{t+1:T}\bigr) 
  \propto f_\theta\bigl(x_{t+1}^{b_{t+1}} \big| x_t^{b_t}\bigr), \label{eq:backwardperfectadaptation}
\end{equation}
where the expression in \eqref{eq:backwardperfectadaptation} is obtained using calculations similar to those in \eqref{eq:CPFBS}.

\subsection{Extension to general auxiliary particle filters}
\label{subsec:apf}
\glsreset{APF}

The previous section demonstrated that the \gls{FAAPF} leads straightforwardly to valid \gls{PMCMC} algorithms and will allow natural connections to be made to certain \gls{EHMM} methods. Here, we show that as was established in \citet[Appendix 8.2]{Pitt2012}, \emph{any} general \emph{\gls{APF}} can be employed in this context and will lead to natural extensions of these methods.

To facilitate later developments, an explicit representation of the associated extended target distribution and related quantities is useful. Viewing the \gls{APF} as a sequential importance resampling algorithm for an appropriate sequence of target distributions as described in \citet{JohansenDoucet2008}, it is immediate that the density associated with such an algorithm is simply:
\begin{align}
  \Psi_\theta^{\mathbf{q}_\theta}(\mathbf{x}_{1:T},\mathbf{a}_{1:T-1}) =
  \prod_{i=1}^N q_{\theta,1}(x_1^i) \prod_{t=2}^T w_{\theta,t-1}^{a_{t-1}^i} q_{\theta,t}\bigl(x_t^i \big|x_{t-1}^{a_{t-1}^i}\bigr),
\end{align}
where $\mathbf{q}_\theta = \{q_{\theta, t}\}_{t=1}^T$ and $q_{\theta,t}$ denotes the proposal distribution employed at time $t$ (with dependence of this distribution upon the observation sequence suppressed from the notation) and $\smash{w_{\theta,t}^i = v_{\theta,t}^i / \sum_{j=1}^N v_{\theta,t}^j}$ with:
\begin{align}
  v_{\theta,t}^i =
  \begin{cases}
    \dfrac{\mu_\theta{\bigl(x_1^i\bigr)} g_\theta{\bigl(y_1|x_1^i\bigr)} \tilde{p}_\theta{\bigl(y_2|x_1^i\bigr)}}{q_{\theta,1}{\bigl(x_1^i\bigr)}}, & \text{if $t = 1$,}\\
    \dfrac{f_\theta{\bigl(x_t^i|x_{t-1}^{a_{t-1}^i}\bigr)} g_\theta{\bigl(y_t|x_t^i\bigr)} \tilde{p}_\theta{\bigl(y_{t+1}|x_t^i\bigr)}}{q_{\theta,t}{\bigl(x_t^i | x_{t-1}^{a_{t-1}^i}\bigr)} \tilde{p}_\theta{\bigl(y_t|x_{t-1}^{a_{t-1}^i}\bigr)}}, & \text{if $1 < t < T$,}\\
    \dfrac{f{\bigl(x_T^i|x_{T-1}^{a_{T-1}^i}\bigr)} g_\theta{\bigl(y_T|x_T^i\bigr)}}{q_{\theta,T}{\bigl(x_T^i|x_{t-1}^{a_{T-1}^i}\bigr)} \tilde{p}_\theta{\bigl(y_T|x_{T-1}^{a_{T-1}^i}\bigr)}}, & \text{if $t = T$,}
  \end{cases}\label{eq:apfweights}
\end{align}
and $\tilde{p}_\theta(y_{t+1}|x_t^i)$ denoting the approximation of the predictive likelihood employed within the weighting of the \gls{APF}. Note that $\tilde{p}_\theta(y_{t+1}|x_t)$ can be any positive function of $x_t$ and the simpler sequential importance resampling \gls{PF} is recovered by setting $\tilde{p}_\theta(y_{t+1}|x_t) \equiv 1$, with the bootstrap \gls{PF} emerging as a particular case thereof when $q_{\theta, t}(x_t|x_{t-1}) = f_\theta(x_t|x_{t-1})$.

Associated with the \gls{APF} is a conditional \gls{PF} of the form:
\begin{equation}
  \phi^{\mathbf{q}_\theta}_\theta{\bigl(\mathbf{x}_{1:T}^{-b_{1:T}},\mathbf{a}_{1:T-1}^{-b_{2:T}}\bigr| x_{1:T}^{b_{1:T}},b_{1:T}\bigr)} 
  = \prod_{\substack{\mathllap{i}=\mathrlap{1}\\\mathllap{i} \neq \mathrlap{b_1}}}^N q_{\theta,1}{\bigl(x_1^i\bigr)} \prod_{t=2}^T \prod_{\substack{\mathllap{i}=\mathrlap{1}\\\mathllap{i} \neq \mathrlap{b_t}}}^N w_{\theta,t-1}^{a_{t-1}^i}\,q_{\theta,t}{\bigl(x_t^i\bigr|x_{t-1}^{a_{t-1}^i}\bigr)}. \label{eq:CAPF}%
\end{equation}
A \gls{PMCMC} algorithm is arrived at by employing the extended target distribution,
\begin{equation}
  \tilde{\pi}^{\mathbf{q}_\theta}{\bigl(\theta,b_{1:T},\mathbf{x}_{1:T},\mathbf{a} _{1:T-1}^{-b_{2:T}}\bigr)}  
  = \frac{1}{N^T} \times \underbrace{\pi{\bigl(  \theta,x_{1:T}^{b_{1:T}}\bigr)}}_{\mathclap{\text{\footnotesize{target}}}} 
  \times  \underbrace{\phi^{\mathbf{q}_\theta}_\theta{\bigl(\mathbf{x}_{1:T}^{-b_{1:T}},\mathbf{a}_{1:T-1}^{-b_{2:T}}\bigr| x_{1:T}^{b_{1:T}},b_{1:T}\bigr)}  }_{\mathclap{\text{\footnotesize{law of conditional \gls{APF}}}}}, \label{eq:APMCMCtarget}%
\end{equation}
and proposal distribution,
\begin{equation}
  q{(\theta,{\theta'})} 
  \times \underbrace{\Psi_{{\theta'}}^{\mathbf{q}_\theta'}{(\mathbf{x}_{1:T},\mathbf{a}_{1:T-1})}}_{\mathclap{\text{\footnotesize{law of \gls{APF}}}}} 
  \times \underbrace{w_{{\theta'},T}^{b_T}}_{\mathclap{\text{\parbox{1.3cm}{\centering\footnotesize{path selection}}}}}. \label{eq:proposalauxiliaryparticlefilter}%
\end{equation}
One can straightforwardly verify that this leads to a \gls{MH} acceptance probability of the form stated in \eqref{eq:PMMHratio} but using the natural unbiased estimator of the normalising constant associated with the \gls{APF},
\begin{equation}
 \hat{p}_\theta(y_{1:T}) = \prod_{t=1}^T \biggl[\frac{1}{N} \sum_{i=1}^N v_{\theta,t}^i\biggr].
\end{equation}
We conclude this section by noting that although the constructions developed above were presented for simplicity with multinomial resampling employed during every iteration of the algorithm, it is straightforward to incorporate more sophisticated, adaptive resampling schemes within this framework.

\section{Original embedded hidden Markov models}
\label{sec:embeddedHMM}

\glsreset{EHMM}
\glsreset{HMM}

\subsection{Extended target distribution}

The \emph{\gls{EHMM}} method of \cite{Neal2003,Neal2004} is based on the introduction of a target distribution on $\varTheta \times \mathcal{X}^{NT}\times \{  1,\dotsc,N \}^N$ of the form
\begin{align}
  \tilde{\pi}(\theta,b_{1:T},\mathbf{x}_{1:T}) 
  = \frac{1}{N^T} \times \underbrace{\pi{\bigl(\theta,x_{1:T}^{b_{1:T}}\bigr)}}_{\mathclap{\text{\footnotesize{target}}}} \times  \underbrace{\prod_{t=1}^T \;\; \Bigl\{ \smashoperator{\prod_{i=b_t -1}^1} \widetilde{R}_{\theta,t}{\bigl(x_t^i\bigr| x_t ^{i+1}\bigr)} \cdot \smashoperator{\prod_{i=b_t+1}^N} R_{\theta,t}{\bigl(x_t^i\bigr| x_t^{i-1}\bigr)}  \Bigr\}}_{\mathclap{\text{\footnotesize{law of conditional random grid generation}}}}, \label{eq:embeddedHMM2003target}
\end{align}
where $R_{\theta,t}$ is a $\rho_{\theta,t}$-invariant Markov transition kernel, i.e.\ $\smash{\int\rho_{\theta,t}(x)  R_{\theta,t}(x^\prime| x) \mathrm{d} x = \rho_{\theta,t}(x^\prime)}$, and $\smash{\widetilde{R}_{\theta,t}}$ is its reversal, i.e.\ $\smash{\widetilde{R}_{\theta,t}(x^\prime|x) = \rho_{\theta,t}(x^\prime) R_{\theta,t}(x|x^\prime) / \rho_{\theta,t}(x)}$ (for $\rho_{\theta,t}$-almost every $x$ and $x^\prime$). 

Similarly to the \gls{PMCMC} extended target distribution, the key feature of $\tilde{\pi}(\theta,b_{1:T},\mathbf{x}_{1:T})$ is that, by construction, it ensures that the associated marginal distribution of $\smash{(\theta,x_{1:T}^{b_{1:T}})}$ is the posterior of interest.

\subsection{Metropolis--Hastings algorithm}
\label{subsec:ehmm-mh}

As detailed in the next section, the algorithm proposed in \citet{Neal2003} can be reinterpreted as a Gibbs sampler targeting $\tilde{\pi}(b_{1:T},\mathbf{x}_{1:T}|\theta)$. We present here an alternative, original \gls{MH} algorithm to sample from $\tilde{\pi}(\theta,b_{1:T},\mathbf{x}_{1:T})$. It relies on a proposal of the form
\begin{equation}
  q{\bigl(  \theta,{\theta'}\bigr)} 
  \times \underbrace{\Psi_{{\theta'}}{\bigl(\mathbf{x}_{1:T}\bigr)}  }_{\mathclap{\parbox{2.3cm}{\text{\parbox{2.3cm}{\centering\footnotesize{law of random grid generation}}}}}} \times \underbrace{q_{{\theta'}}{\bigl(b_{1:T}\bigr| \mathbf{x}_{1:T}\bigr)}}_{\mathclap{\text{\parbox{2cm}{\centering\footnotesize{path selection}}}}}, \label{eq:proposalembeddedHMM2003}
\end{equation}
where
\begin{equation}
  \Psi_\theta{\bigl(  \mathbf{x}_{1:T}\bigr)} 
  \coloneqq \frac{1}{N^T} \prod_{t=1}^T \Bigl\{ \rho_{\theta,t}{\bigl(x_t^1\bigr)} \smashoperator{\prod_{i=2}^N} R_{\theta,t}{\bigl(x_t^i\bigr| x_t^{i-1}\bigr)} \Bigr\} \label{eq:distributionrandomHMMfilter}
\end{equation}
is sometimes referred to as the \emph{ensemble base measure} \citep{Neal2011} and
\begin{equation}
  q_\theta(b_{1:T} | \mathbf{x}_{1:T})
  \coloneqq \frac{\tilde{p}_\theta{\bigl(  x_{1:T}^{b_{1:T}},y_{1:T}\bigr)}}{\sum_{b_{1:T}^\prime}\tilde{p}_\theta{\bigl(  x_{1:T}^{b_{1:T}^\prime},y_{1:T}\bigr)}}
  = \frac{1}{N^T}\frac{\tilde{p}_\theta{\bigl(x_{1:T}^{b_{1:T}},y_{1:T}\bigr)}}{\tilde{p}_\theta(y_{1:T})}. \label{eq:pathselectionEHMM}%
\end{equation}
In this expression, we have (where we note that this is no longer a probability density with respect to Lebesgue measure)%
\begin{equation}
  \tilde{p}_\theta(  x_{1:T},y_{1:T}) 
  \coloneqq \frac{\mu_\theta(x_1) g_\theta(y_1|x_1)}{\rho_{\theta,1}(x_1)} \prod_{t=2}^T \frac{f_\theta(x_t|x_{t-1}) g_\theta(y_t|x_t)}{\rho_{\theta,t}(x_t)} \label{eq:modifiedjointposterior}%
\end{equation}
and
\begin{equation}
\tilde{p}_\theta(y_{1:T})
\coloneqq \frac{1}{N^T}\sum
_{b_{1:T}^\prime}\tilde{p}_\theta{\bigl(  x_{1:T}^{b_{1:T}^\prime%
},y_{1:T}\bigr)}  . \label{eq:likelihoodestimatorembeddedHMM}%
\end{equation}
To sample from $\Psi_\theta( \mathbf{x}_{1:T})$, we sample $\smash{x_t^{1}\sim\rho_{\theta,t}}(x_t^{1})$ and $\smash{\mathbf{x}_t^{-1}}\sim\smash{\prod_{i=2}^N R_{\theta,t}(x_t^i| x_t^{i-1})}$ for $t=1,...,T$. Hence, at time $t$ all of the particles are marginally distributed according to $\rho_{\theta,t}$. When  $\smash{R_{\theta,t}(x^\prime| x) = \rho_{\theta,t}(x^\prime)}$, this corresponds to the algorithm proposed in \citet{LinChen2005}. Sampling from the high-dimensional discrete distribution $q_\theta(b_{1:T}| \mathbf{x}_{1:T})$ can be performed in $O(N^2T)$ operations with the finite state-space \gls{HMM} filter using the $N$ states $(x_t^i)$ at time $t$, transition probabilities proportional to $f_\theta(x_t^j| x_{t-1}^i)$ and conditional probabilities of the observations proportional to $g_\theta(y_t| x_t^i) / \rho_{\theta,t}(x_t^i)$. We also obtain as a by-product $\tilde{p}_\theta(y_{1:T})$, which is an unbiased estimate of $p_\theta(y_{1:T})$.

The resulting \gls{MH} algorithm targeting the extended distribution given in \eqref{eq:embeddedHMM2003target} with the proposal given in \eqref{eq:proposalembeddedHMM2003} admits an acceptance probability of the form
\begin{equation}
  1 \wedge 
  \frac{\tilde{p}_{{\theta'}}(y_{1:T})p({\theta'})}{\tilde{p}_\theta(y_{1:T})
  p(\theta)}
  \frac{q({\theta'},\theta)}{q(\theta,{\theta'})},
  \label{eq:embeddedHMMratio}%
\end{equation}
i.e.\ it looks very much like the \gls{PMMH} algorithm, except that instead of having likelihood terms estimated by a particle filter, these likelihood terms are estimated using a finite state-space \gls{HMM} filter.

To establish the correctness of the acceptance probability given in \eqref{eq:embeddedHMMratio}, we note that
\begin{align}
  \frac{\tilde{\pi}(\theta,b_{1:T},\mathbf{x}_{1:T})}{\Psi_\theta(\mathbf{x}_{1:T}) q_\theta(b_{1:T}| \mathbf{x}_{1:T}) }
  &  = \frac{N^{-T}\pi(\theta,x_{1:T}^{b_{1:T}}) \prod_{t=1}^T
  \bigl\{ \prod_{i=b_t -1}^1 \widetilde{R}_{\theta,t}(x_t^i| x_t^{i+1}) \cdot \prod_{i=b_t+1}^N
  R_{\theta,t}(x_t^i| x_t^{i-1})  \bigr\} }{
  \prod_{t=1}^T \bigl\{ \rho_{\theta,t}(  x_t^{1}) \cdot \prod_{i=2}^N R_{\theta,t}(x_t^i| x_t^{i-1})  \bigr\}
  N^{-T}\frac{\tilde{p}_\theta(  x_{1:T}^{b_{1:T}},y_{1:T}) }{\tilde{p}_\theta(  y_{1:T})  }}\\
  &  =\frac{p_\theta(  x_{1:T}^{b_{1:T}},y_{1:T})  / p_\theta(y_{1:T})  }{\prod_{t=1}^T \rho_{\theta,t}(x_t^{b_t})} 
  \biggl[ \frac{p_\theta(x_{1:T}^{b_{1:T}},y_{1:T})}{\tilde{p}_\theta(y_{1:T}) \prod_{t=1}^T \rho_{\theta,t}(  x_t^{b_t})}\biggr]^{-1}\\
  & = p(\theta| y_{1:T}) \frac{\tilde{p}_\theta(y_{1:T})}{p_\theta(y_{1:T})} \propto \tilde{p}_\theta(y_{1:T})  p(\theta), \label{eq:embeddedHMMratiotargetproposal}
\end{align}
where we have used that 
\begin{equation}
  \frac{\tilde{p}_\theta(x_{1:T},y_{1:T})}{\tilde{p}_\theta{\bigl(  y_{1:T}\bigr)}}
  =\frac{p_\theta(x_{1:T},y_{1:T})}{\tilde{p}_\theta(y_{1:T}) \prod_{t=1}^T \rho_{\theta,t}{\bigl(x_t^{b_t}\bigr)}}.
\end{equation}
In addition, we have used the following identity which we will also exploit in the next section: if $R$ is a $\rho$-invariant Markov kernel and $\widetilde{R}$ the associated reversal, then for any $b,c \in \{1,\dotsc,N\}$,
\begin{align}
 \smashoperator{\prod_{i=b-1}^1} \widetilde{R}{\bigl(x^i\bigr| x^{i+1}\bigr)} \cdot \rho{\bigl(x^{b}\bigr)} \cdot \smashoperator{\prod_{i=b+1}^N} R{\bigl(x^i\bigr| x^{i-1}\bigr)}  = \smashoperator{\prod_{i=c-1}^1} \widetilde{R}{\bigl(x^i\bigr| x^{i+1}\bigr)} \cdot \rho{\bigl(x^c\bigr)} \cdot \smashoperator{\prod_{i=c+1}^N} R{\bigl(x^i\bigr| x^{i-1}\bigr)}. \label{eq:useful_reversal_kernel_identity}
\end{align}

\subsection{Interpretation as a collapsed Gibbs sampler}
\label{subsec:ehmm-gibbs}

Consider the following Gibbs sampler type algorithm to sample from $\pi(x_{1:T}|\theta)$: 
\begin{enumerate}
 \item \label{enum:ehmm_gibbs:1} Sample $b_{1:T}$ uniformly at random on $\smash{\{1,\dotsc,N\}^T}$ and set $\smash{x_{1:T}^{b_{1:T}}\leftarrow x_{1:T}}$;
 \item \label{enum:ehmm_gibbs:2} Sample $\smash{\tilde{\pi}(\mathbf{x}_{1:T}^{-b_{1:T}}| \theta,b_{1:T},x_{1:T}^{b_{1:T}})}$;
 \item \label{enum:ehmm_gibbs:3} Sample $\smash{b_T\sim\tilde{\pi}(b_T|\theta,\mathbf{x}_{1:T})}$ then $\smash{b_{T-1}\sim\tilde{\pi}(b_{T-1}| \theta,\mathbf{x}_{1:T-1},x_T^{b_T},b_T)}$ and so on.
\end{enumerate}
It is obvious that Steps~\ref{enum:ehmm_gibbs:1} and \ref{enum:ehmm_gibbs:2} coincide with the first steps of the \gls{EHMM} algorithm described in \citet{Neal2003}. For Step~\ref{enum:ehmm_gibbs:3}, we note that
\begin{align}
  \MoveEqLeft \tilde{\pi}{\bigl(b_t\bigr| \theta,\mathbf{x}_{1:t},x_{t+1:T}^{b_{t+1:T}},b_{t+1:T}\bigr)} \\
  & \propto \smashoperator{\sum_{b_{1:t-1}}} \idotsint
  \pi{\bigl(\theta,x_{1:T}^{b_{1:T}}\bigr)}
  \prod_{n=1}^T \;\;
  \Bigl\{
  \smashoperator{\prod_{i=b_n -1}^1}
  \widetilde{R}_{\theta,n}{\bigl(x_n^i\bigr| x_n^{i+1}\bigr)} \cdot
  \smashoperator{\prod_{i=b_n+1}^N} R_{\theta,n}{\bigl(x_n^i\bigr| x_n^{i-1}\bigr)} \Bigr\}\,
  \mathrm{d}\mathbf{x}_{n+1}^{-b_{n+1}}\cdots \mathrm{d} \mathbf{x}_T^{-b_T}\\
  & =\smashoperator{\sum_{b_{1:t-1}}} \pi{\bigl(\theta,x_{1:T}^{b_{1:T}}\bigr)}
  \smashoperator{\prod_{n=1}^t} \;\;
  \Bigl\{\smashoperator{\prod_{i=b_n -1}^1}
  \widetilde{R}_{\theta,n}{\bigl(x_n^i \bigr| x_n^{i+1}\bigr)} \cdot \smashoperator{\prod_{i=b_n+1}^N}
  R_{\theta,n}{\bigl(x_n^i\bigr| x_n^{i-1}\bigr)}  \Bigr\}\\
  &  = \smashoperator{\sum_{b_{1:t-1}}} \pi{\bigl(\theta,x_{1:T}^{b_{1:T}}\bigr)} \prod_{n=1}^t  \frac{
  \prod_{i=b_n -1}^1 \widetilde{R}_{\theta,n}{\bigl(x_n^i\bigr| x_n^{i+1}\bigr)} \cdot \rho_{\theta,n}{\bigl(  x_n^{b_n}\bigr)} \cdot \prod_{i=b_n+1}^N R_{\theta,n}{\bigl(x_n^i\bigr| x_n^{i-1}\bigr)}
  }{\rho_{\theta,n}{\bigl(  x_n^{b_n}\bigr)}  }\\
  &  \propto\smashoperator{\sum_{b_{1:t-1}}}\,\frac{\pi{\bigl(  \theta,x_{1:T}^{b_{1:T}}\bigr)}  }{
  \prod_{n=1}^t \rho_{\theta,n}{\bigl(  x_n^{b_n}\bigr)} },
\end{align}
where by \eqref{eq:useful_reversal_kernel_identity}, the numerator in the penultimate line is independent of $b_n$. Since
\begin{align}
  \frac{\pi{\bigl(\theta,x_{1:T}^{b_{1:T}}\bigr)}}{\prod_{n=1}^t \rho_{\theta,n}{\bigl(x_n^{b_n}\bigr)}}  
  & \propto \frac{p_{\theta
  }{\bigl(  x_{1:T}^{b_{1:T}},y_{1:T}\bigr)} }{\prod_{n=1}^t \rho_{\theta,n}{\bigl(x_n^{b_n}\bigr)}}\\
  & \propto 
   \underbrace{\prod_{n=1}^t \frac{f_\theta\bigl(x_n^{b_n}\bigr| x_{n-1}^{b_{n-1}}\bigr)g_\theta{\bigl(y_n\bigr| x_n^{b_n}\bigr)}}{\rho_{\theta,n}{\bigl(x_n^{b_n}\bigr)}  }}_{\mathclap{\text{\footnotesize{modified posterior $\tilde{p}_\theta(x_{1:t}^{b_{1:t}}| y_{1:t})$}}}} \cdot \smashoperator{\prod_{n=t+1}^T} f_\theta{\bigl(x_n^{b_n}\bigr| x_{n-1}^{b_{n-1}}\bigr)} g_\theta{\bigl(y_n\bigr| x_n^{b_n}\bigr)},
\end{align}
we can compute the marginal $\smash{\tilde{p}_\theta(x_t^{b_t}| y_{1:t}) \coloneqq \sum_{b_{1:t-1}} \tilde{p}_\theta(x_{1:t}^{b_{1:t}}| y_{1:t})}$ using the same (finite state-space) \gls{HMM} filter discussed in the previous section and so
\begin{equation}
 \tilde{\pi}{\bigl(b_t\bigr| \theta,\mathbf{x}_{1:t},x_{t+1:T}^{b_{t+1:T}},b_{t+1:T}\bigr)}  \propto\tilde{p}_\theta{\bigl(x_t^{b_t}\bigr| y_{1:t}\bigr)}  f_\theta{\bigl(x_{t+1}^{b_t+1}\bigr| x_t^{b_t}\bigr)}
\end{equation}
coinciding with the expression obtained in \citet{Neal2003}. This is an alternative proof of validity of the algorithm. The present derivation is more complex than that in \citet{Neal2003} which relies on a simple detailed balance argument. One potential benefit of our approach is that it can be extended systematically to any extended target admitting a similar structure; see for example \citet[p.~116]{LindstenSchon2013} for extensions to the non-Markovian case. Finally, we note that this algorithm may be viewed as a special case of the framework proposed in \citet{Tjelmeland2004} and simplifies to Barker's kernel \citep{Barker1965} if $N=2$ and $T=1$.

\section{Alternative embedded hidden Markov models}
\label{sec:embeddedHMMnewversion}

\glsreset{MCMCPF}
\glsreset{MCMCFAAPF}
\glsreset{MCMCAPF}

In its original version, the \gls{EHMM} method has a computational cost per iteration of order  $O(N^2T)$  compared to $O(NT)$ for \gls{PMCMC} methods and it samples particles independently across time which can be inefficient if the latent states are strongly correlated. The new version of \gls{EHMM} methods, which was proposed in \citet{ShestopaloffNeal2016}, resolves both of these limitations. It  can be viewed as a \gls{PMCMC}-type algorithm making use of a new type of \gls{PF} that we term the \emph{\gls{MCMCFAAPF}} given its connection to the \gls{FAAPF} which we detail below.

\subsection{Extended target distribution}

This version of the \gls{EHMM}, henceforth referred to as the \emph{alternative} \gls{EHMM} method,  relies on the extended target distribution
\begin{equation}
 \tilde{\pi}{\bigl(  \theta,b_{1:T},\mathbf{x}_{1:T}, \mathbf{a}_{1:T-1}^{-b_{2:T}}\bigr)}  = \frac{1}{N^T} \times \underbrace{\pi{\bigl(\theta,x_{1:T}^{b_{1:T}}\bigr)}}_{\mathclap{\text{\footnotesize{target}}}} \times \underbrace{\phi_\theta\bigl(\mathbf{x}_{1:T}^{-b_{1:T}},\mathbf{a}_{1:T-1}^{-b_{2:T}}\bigr|x_{1:T}^{b_{1:T}},b_{1:T}\bigr)}_{\mathclap{\text{\footnotesize{law of conditional \gls{MCMCFAAPF}}}}},
\end{equation}
where we will refer to the algorithm inducing the following distribution as the conditional \gls{MCMCFAAPF} for reasons which are made clear below:
\begin{align}
  \MoveEqLeft \phi_\theta{\bigl( \mathbf{x}_{1:T}^{-b_{1:T}},\mathbf{a}_{1:T-1}^{-b_{2:T}}\bigr| x_{1:T}^{b_{1:T}},b_{1:T}\bigr)}\\*
  & = 
  \smashoperator{\prod_{\smash{i=b_1-1}}^1}\widetilde{R}_{\theta,1}{\bigl(x_1^i\bigr| x_1^{i+1}\bigr)}  \cdot \smashoperator{\prod_{\smash{i=b_1+1}}^N} R_{\theta,1}{\bigl(x_1^i\bigr| x_1^{i-1}\bigr)} \label{eq:conditionalPFnovelEHMMtarget}\\
  &  \quad \times \smashoperator{\prod_{t=2}^{\smash{T}}} \;\; \Bigl\{\smashoperator{\prod_{i=b_t -1}^{\smash{1}}} \widetilde{R}_{\theta,t}{\bigl(x_t^i,a_{t-1}^i\bigr| x_t^{i+1},a_{t-1}^{i+1};\mathbf{x}_{t-1}\bigr)} \smashoperator{\prod_{i=b_t+1}^{\smash{N}}} R_{\theta,t}{\bigl(x_t^{i+1},a_{t-1}^{i+1}\bigr| x_t^i,a_{t-1}^i;\mathbf{x}_{t-1}\bigr)}  \Bigr\},
\end{align}
with $\smash{b_t=a_t^{b_{t+1}}}$ as for \gls{PMCMC} methods.

Here $R_{\theta,1}$ is invariant with respect to $\rho_{\theta,1}(x_1) = p_\theta(x_1 | y_1)$ whereas, for $t=2,\dotsc,T$,  $R_{\theta,t}(\,\cdot\,|\,\cdot\,; \mathbf{x}_{t-1})$ is invariant w.r.t.\
\begin{align}
  \rho_{\theta,t}{\bigl(x_t, a_{t-1}\bigr| \mathbf{x}_{t-1}\bigr)} 
  & =\frac{g_\theta{\bigl(y_t\bigr| x_t\bigr)} f_\theta{\bigl(x_t\bigr| x_{t-1}^{a_{t-1}}\bigr)}} {\sum_{i=1}^Np_\theta{\bigl(y_t\bigr| x_{t-1}^i\bigr)}}
  =\frac{p_\theta{\bigl(y_t\bigr| x_{t-1}^{a_{t-1}}\bigr)}}{\sum_{i=1}^Np_\theta{\bigl(y_t\bigr| x_{t-1}^i\bigr)}}p_\theta{\bigl(x_t\bigr| y_t,x_{t-1}^{a_{t-1}}\bigr)}  ,
\end{align}
while, for $t=1,\dotsc,T$, $\widetilde{R}_{\theta,t}(\,\cdot\,|\,\cdot\,; \mathbf{x}_{t-1})$ denotes the reversal of the kernel $R_{\theta,t}(\,\cdot\,|\,\cdot\,; \mathbf{x}_{t-1})$ with respect to its invariant distribution.

Note that if $R_{\theta,1}(x_1^\prime|x_1)  =\rho_{\theta,1}(  x_1^\prime)$ and $R_{\theta,t}(x_t^\prime, a_{t-1}^\prime|x_t, a_{t-1};\mathbf{x}_{t-1}) = \rho_{\theta,t}(x_t^\prime, a_{t-1}^\prime| \mathbf{x}_{t-1})$, the extended target $\tilde{\pi}( \theta,b_{1:T},\mathbf{a}_{1:T-1}^{-b_{2:T}}, \mathbf{x}_{1:T})$ coincides exactly with the extended target associated with the \gls{FAAPF} described in Section~\ref{Section:perfectadaptationPF}. As explored in the following two sections, this allows us to understand this \gls{EHMM} approach as the incorporation of a slightly more general class of \glspl{PF} within a \gls{PMCMC} framework and ultimately suggests further generalisations of these algorithms.

\subsection{Metropolis--Hastings algorithm}
\label{subsec:mcmc-fa-apf}

We now consider the following \gls{MH} algorithm to sample from $\smash{\tilde{\pi}(\theta,b_{1:T}, \mathbf{x}_{1:T}, \mathbf{a}_{1:T-1}^{-b_{2:T}})}$. It relies on a proposal of the form
\begin{equation}
  q{\bigl(  \theta,{\theta'}\bigr)} \times \underbrace{\Psi_{{\theta'}}{\bigl(\mathbf{x}_{1:T}, \mathbf{a}_{1:T-1}\bigr)}  }_{\mathclap{\text{\parbox{2cm}{\centering\footnotesize{law of \gls{MCMCFAAPF}}}}}} \times \underbrace{\frac{1}{N}}_{\mathclap{\text{\parbox{1.3cm}{\centering\footnotesize{path selection}}}}}, \label{eq:proposalnovelEHMM}
\end{equation}
i.e.\ to sample $b_{1:T}$, we pick $b_T$ uniformly at random, then set $\smash{b_t=a_t^{b_{t+1}}}$ for $t=T-1,\dotsc,1$. Moreover,
\begin{align}
  \Psi_\theta{\bigl( \mathbf{x}_{1:T}, \mathbf{a}_{1:T-1}\bigr)} 
  & = \rho_{\theta,1}{\bigl(  x_1^1\bigr)} \smashoperator{\prod_{\smash{i=2}}^N} R_{\theta,1}{\bigl(x_1^i\bigr| x_1^{i-1}\bigr)}\\
  &  \quad \times \smashoperator{\prod_{t=2}^{\smash{T}}} \; \Bigl\{
  \rho_{\theta,t}{\bigl( x_t^1, a_{t-1}^1\bigr| \mathbf{x}_{t-1}\bigr)} \smashoperator{\prod_{i=2}^{\smash{N}}} R_{\theta,t}{\bigl(x_t^{i},a_{t-1}^{i}\bigr| x_t^{i-1},a_{t-1}^{i-1};\mathbf{x}_{t-1}\bigr)}  \Bigr\}   \label{eq:novelPFforEHMM}
\end{align}
is the law of a novel \gls{PF} type algorithm, which we refer to as the \gls{MCMCFAAPF}; again the reason for this terminology should become clear below.

The \gls{MCMCFAAPF} proceeds as follows. 
\begin{enumerate}
 \item At time $1$, sample $x_1^1\sim\rho_{\theta,1}(x_1^1)$ and then $\mathbf{x}_1^{-1}\sim \prod_{i=2}^N R_{\theta,1}(x_1^i| x_1^{i-1})$. 
 \item At time $t=2,\dotsc,T$, sample
 \begin{enumerate}
  \item $(x_t^1, a_{t-1}^1) \sim\rho_{\theta,t}(x_t^1, a_{t-1}^1| \mathbf{x}_{t-1})$,
  \item $(\mathbf{x}_t^{-1},\mathbf{a}_{t-1}^{-1})  \sim \prod_{i=2}^N R_{\theta,t}(x_t^{i},a_{t-1}^{i}| x_t^{i-1},a_{t-1}^{i-1};\mathbf{x}_{t-1})$. 
 \end{enumerate}
\end{enumerate}
If $R_{\theta,1}(x_1^\prime\bigr| x_1) = \rho_{\theta,1}(x_1^\prime)$ and $R_{\theta,t}(x_t^\prime, a_{t-1}^\prime\bigr|x_t, a_{t-1};\mathbf{x}_{t-1}) = \rho_{\theta,t}(x_t^\prime, a_{t-1}^\prime\bigr| \mathbf{x}_{t-1})$, this corresponds to the standard \gls{FAAPF}.

The resulting \gls{MH} algorithm targeting the extended distribution defined in \eqref{eq:embeddedHMM2003target} and using the proposal defined in \eqref{eq:proposalembeddedHMM2003} admits an acceptance probability of the form
\begin{equation}
  1 \wedge \frac{\hat{p}_{{\theta'}}(y_{1:T}) p({\theta'})}{\hat{p}_\theta(y_{1:T}) p(\theta)}\frac{q({\theta'},\theta)}{q(\theta,{\theta'})}, \label{eq:alternative_ehmm_acceptance_probability}
\end{equation}
i.e.\ it looks very much like the \gls{PMMH}, except that here $\hat{p}_\theta(y_{1:T})$ is given by the expression in \eqref{eq:likelihoodestimatorperfectadaptationPF} with particles generated via \eqref{eq:novelPFforEHMM}. Note that this estimate is unbiased.

The validity of the acceptance probability in \eqref{eq:alternative_ehmm_acceptance_probability} can be established by calculating
\begin{align}
  \MoveEqLeft \frac{\tilde{\pi}{\bigl(  \theta,b_{1:T}, \mathbf{x}_{1:T}, \mathbf{a}_{1:T-1}^{-b_{2:T}}\bigr)}}{\Psi_\theta{\bigl(  \mathbf{a}_{1:T-1},\mathbf{x}_{1:T}\bigr)}  \frac{1}{N}}\\
  & = N\pi{\bigl(\theta,x_{1:T}^{b_{1:T}}\bigr)}  \frac{\prod_{i=b_1-1}^1 \widetilde{R}_{\theta,1}{\bigl(x_1^i\bigr| x_1^{i+1}\bigr)}  \cdot \prod_{i=b_1+1}^N R_{\theta,1}{\bigl(x_1^i\bigr| x_1^{i-1}\bigr)}}{\rho_{\theta,1}{\bigl(  x_1^1\bigr)}  \prod_{i=2}^N R_{\theta,1}{\bigl(x_1^j\bigr| x_1^{i-1}\bigr)}}\\
  & \quad \times \prod_{t=2}^T \frac{\prod_{i=b_t -1}^1
  \widetilde{R}_{\theta,t}{\bigl(x_t^i,a_{t-1}^i\bigr|x_t^{i+1},a_{t-1}^{i+1};\mathbf{x}_{t-1}\bigr)}  \cdot \prod_{i=b_t+1}^N R_{\theta,t}{\bigl(x_t^{i+1},a_{t-1}^{i+1}\bigr| x_t^i,a_{t-1}^i;\mathbf{x}_{t-1}\bigr)}}{\rho_{\theta,t}{\bigl(x_t^{1}, a_{t-1}^{1}\bigr| \mathbf{x}_{1:t-1}\bigr)} \prod_{i=2}^N R_{\theta,t}{\bigl(x_t^{i+1},a_{t-1}^{i+1}\bigr| x_t^i,a_{t-1}^j;\mathbf{x}_{t-1}\bigr)}}\\
  & = \frac{N^{T-1}\pi{\bigl(  \theta,x_{1:T}^{b_{1:T}}\bigr)}  }{\rho_{\theta,1}{\bigl(  x_1^{b_1}\bigr)}
  \prod_{t=2}^T \rho_{\theta,t}{\bigl(x_t^{b_t}, a_{t-1}^{b_t}\bigr|\mathbf{x}_{1:t-1}\bigr)}  }\\
  & = \frac{N^{T-1}\pi{\bigl(\theta,x_{1:T}^{b_{1:T}}\bigr)}}{\frac{p_\theta(x_1^{b_1}, y_1)}{p_\theta(y_1)} \prod_{t=2}^T \frac{f_\theta(x_t^{b_t}| x_{t-1}^{b_{t-1}}) g(y_t| x_t^{b_t})}{\sum_{i=1}^N p_\theta(y_t| x_{t-1}^i)  } }
  = p(\theta | y_{1:T})  \frac{\hat{p}_\theta(y_{1:T})}{p_\theta(y_{1:T})}.
\end{align}
We have again used identity \eqref{eq:useful_reversal_kernel_identity} and additionally that $\smash{b_t=a_t^{b_{t+1}}}$, for $t = T-1,\dotsc, 1$.

\subsection{Gibbs sampler}
\label{subsec:gibbs_mcmc-fa-apf}

The \gls{EHMM} method of \cite{ShestopaloffNeal2016} can be reinterpreted as a collapsed Gibbs sampler to sample from the extended target distribution $\tilde{\pi}(\theta,b_{1:T},\mathbf{x}_{1:T}, \mathbf{a}_{1:T-1}^{-b_{2:T}})$. Given a current value of $x_{1:T}$, the algorithm proceeds as follows.
\begin{enumerate}
 \item Sample $b_{1:T}$ uniformly at random and set $x_{1:T}^{b_{1:T}}\leftarrow x_{1:T}$.
 \item Run the conditional \gls{MCMCFAAPF}, i.e.\ sample from $\phi_\theta(\mathbf{x}_{1:T}^{-b_{1:T}},\mathbf{a}_{1:T-1}^{-b_{2:T}}| x_{1:T}^{b_{1:T}},b_{1:T})$.
 \item Sample $b_T$ according to $\Pr(b_T=m) = 1/N$ and then, for $t=T-1,\dotsc,1$, sample $b_t$ according to a distribution proportional to $f_\theta(x_{t+1}^{b_{t+1}}| x_t^{b_t})$. 
\end{enumerate}
The validity of the algorithm is established using a detailed balance argument in \citet{ShestopaloffNeal2016}. Alternatively, we can show using simple calculations similar to the ones presented earlier that
\begin{equation}
 \tilde{\pi}\bigl( b_t\big| \theta,\mathbf{x}_{1:t}, x_{t+1:T}^{b_{t+1:T}}, b_{t+1:T}\bigr)
 \propto f_\theta\bigl(x_{t+1}^{b_{t+1}} \big| x_t^{b_t}\bigr)  .
\end{equation}
In the standard conditional \gls{PF}, the particles are conditionally independent given the previously sampled values. The conditional \gls{MCMCFAAPF} allows for conditional dependence between all the particles (and ancestor indices) generated in one time step. Indeed, we can choose the kernels $R_{\theta, t}^{\mathbf{q}_\theta}(\,\cdot\,|\,\cdot\,;\mathbf{x}_{t-2:t-1}, \mathbf{a}_{t-2})$ such that they induce only small, local moves. This can improve the performance of \gls{PG} samplers in high dimensions: as with standard \gls{MCMC} schemes, less ambitious local moves are much more likely to be accepted. Of course, as with any local proposal one could not expect such a strategy to work well with strongly multi-modal target distributions without further refinements.

\section{Novel practical extensions}
\label{sec:novel_methodology}

Motivated by the connections identified above, we now develop extensions based upon the more general \gls{PMCMC} algorithms described above, in particular considering constructions based around general \glspl{APF}. In particular, we relax the requirement in the \gls{MH} algorithm from Section~\ref{subsec:mcmc-fa-apf} that it is possible to sample from the proposal distribution of the \gls{FAAPF} (which is possible in only a small number of tractable models) and to compute its associated importance weight.

\subsection{MCMC APF}

Generalising the \gls{MCMCFAAPF} in the same manner as the \gls{APF} generalises the \gls{FAAPF} leads us to propose a (general) \emph{\gls{MCMCAPF}}. Set
\begin{align}
  \rho_{\theta,t}^{\mathbf{q}_\theta}(x_t,a_{t-1}|\mathbf{x}_{t-2:t-1},  \mathbf{a}_{t-2}) 
  & = 
  \begin{cases}
    q_{\theta,1}(x_1), & \text{if $t = 1$,}\\
    \dfrac{v_{\theta,t-1}^{a_{t-1}}}{\sum_{i=1}^N v_{\theta,t-1}^i} q_{\theta,t}(x_t|x_{t-1}^{a_{t-1}}), & \text{if $t> 1$,}
  \end{cases}
\end{align}
where $v_{\theta,t-1}^i$ are as defined in \eqref{eq:apfweights}, and is responsible for the dependence upon $a_{t-1}$ and $x_{t-2}$ in particular, and we allow $\smash{R_t^{\mathbf{q}_\theta}(\,\cdot\,|\,\cdot\,;\mathbf{x}_{t-2:t-1}, \mathbf{a}_{t-2})}$ and $\smash{\widetilde{R}_{t}^{\mathbf{q}_\theta}(\,\cdot\,|\,\cdot\,;\mathbf{x}_{t-2:t-1}, \mathbf{a}_{t-2})}$ to respectively denote a $\rho_{\theta,t}^{\mathbf{q}_\theta}(\,\cdot\,|\mathbf{x}_{t-2:t-1}, \mathbf{a}_{t-2})$-invariant Markov kernel and the associated reversal kernel. Although this expression superficially resembles the mixture proposal of the \emph{marginalised} \gls{APF} \citep{Klass2005}, by explicitly including the ancestry variables it avoids incurring the $O(N^2)$ cost and allows an approximation of smoothing distributions. We then define the law of the \gls{MCMCAPF} via:
\begin{align}
  \Psi^{\mathbf{q}_\theta}_\theta{\bigl(\mathbf{x}_{1:T}, \mathbf{a}_{1:T-1}\bigr)}   
   & \coloneqq \rho^{\mathbf{q}_\theta}_{\theta,1}{\bigl(  x_1^1\bigr)}  
  \smashoperator{\prod_{\smash{i=2}}^N} R^{\mathbf{q}_\theta}_{\theta,1}{\bigl(x_1^i\bigr| x_1^{i-1}\bigr)}\\
  & \quad \times
  \smashoperator{\prod_{t=2}^{\smash{T}}}\;\Bigl\{\rho^{\mathbf{q}_\theta}_{\theta,t}{\bigl(x_t^1, a_{t-1}^1\bigr| \mathbf{x}_{t-2:t-1}, \mathbf{a}_{t-2}\bigr)}
  \bigr. \smashoperator{\prod_{i=2}^{\smash{N}}}
  R^{\mathbf{q}_\theta}_{\theta,t}{\bigl(x_t^{i},a_{t-1}^{i}\bigr| x_t^{i-1},a_{t-1}^{i-1};\mathbf{x}_{t-2:t-1}, \mathbf{a}_{t-2}\bigr)} \Bigr\}.
\end{align}
The corresponding extended \gls{PMCMC} target distribution is simply:
\begin{equation}
  \tilde{\pi}^{\mathbf{q}_\theta}{\bigl(  \theta,b_{1:T},\mathbf{x}_{1:T}, \mathbf{a}_{1:T-1}^{-b_{2:T}}\bigr)} 
  = \frac{1}{N^T} \times \underbrace{\pi\bigl(\theta,x_{1:T}^{b_{1:T}}\bigr)  }_{\text{\footnotesize{target}}} \times \underbrace{\phi^{\mathbf{q}_\theta}_{\theta}\bigl( \mathbf{x}_{1:T}^{-b_{1:T}},\mathbf{a}_{1:T-1}^{-b_{2:T}}\big\vert x_{1:T}^{b_{1:T}},b_{1:T}\bigr)}_{\text{\footnotesize{law of conditional \gls{MCMCAPF}}}}, \label{eq:novelEHMMtarget}
\end{equation}
where, as might be expected:
\begin{align}
  \MoveEqLeft \phi^{\mathbf{q}_\theta}_\theta{\bigl(\mathbf{x}_{1:T}^{-b_{1:T}},\mathbf{a}_{1:T-1}^{-b_{2:T}}\bigr| x_{1:T}^{b_{1:T}},b_{1:T}\bigr)}\\*
   &  = \smashoperator{\prod_{\smash{i=b_1-1}}^1}
  \widetilde{R}^{\mathbf{q}_\theta}_{\theta,1}{\bigl(x_1^i\bigr| x_1^{i+1}\bigr)}  \cdot \smashoperator{\prod_{\smash{i=b_1+1}}^N} R^{\mathbf{q}_\theta}_{\theta,1}{\bigl(x_1^i\bigr| x_1^{i-1}\bigr)}\\
  & \quad \times \smashoperator{\prod_{t=2}^{\smash{T}}} \;\; \Bigl\{\smashoperator{\prod_{i=b_t-1}^{\smash{1}}} \widetilde{R}^{\mathbf{q}_\theta}_{\theta,t}{\bigl(x_t^i,a_{t-1}^i\bigr| x_t^{i+1},a_{t-1}^{i+1};\mathbf{x}_{t-2:t-1}, \mathbf{a}_{t-2}\bigr)}  \bigr. \smashoperator{\prod_{i=b_t+1}^{\smash{N}}} R^{\mathbf{q}_\theta}_{\theta,t}{\bigl(x_t^{i},a_{t-1}^{i}\bigr| x_t^{i-1},a_{t-1}^{i-1};\mathbf{x}_{t-2:t-1}, \mathbf{a}_{t-2}\bigr)}  \Bigr\}. 
\end{align}

Note that the \gls{MCMCFAAPF} can be viewed as a special case of the \gls{MCMCAPF} in much the same way that the \gls{FAAPF} from Section~\ref{Section:perfectadaptationPF} can be viewed as a special case of the (general) \gls{APF} from Section~\ref{subsec:apf}. 
  
\subsection{Metropolis--Hastings algorithms}

We arrive at a \gls{PMMH}-type algorithm based around the \gls{MCMCAPF} by considering proposal distributions of the form:
\begin{equation}
  q{\bigl(\theta,{\theta'}\bigr)} \times \underbrace{\Psi^{\mathbf{q}_\theta'}_{{\theta'}}{\bigl(\mathbf{x}_{1:T}, \mathbf{a}_{1:T-1}\bigr)}}_{\mathclap{\text{\parbox{2.3cm}{\centering\footnotesize{law of \gls{MCMCAPF}}}}}} \times \underbrace{w_{{\theta'},T}^{b_T}}_{\mathclap{\text{\parbox{1.3cm}{\centering\footnotesize{path selection}}}}}, 
\end{equation}
where, as in Section~\ref{subsec:apf}, $\smash{w_{\theta,T}^i = v_{\theta,T}^i / \sum_{j=1}^N v_{\theta,T}^j}$ and $\smash{\hat{p}_\theta(y_{1:T}) = \prod_{t=1}^T N^{-1} \sum_{i=1}^N v_{\theta,t}^i}$ is again an unbiased estimate of the marginal likelihood. 

Note that the \gls{PMMH}-type variant of the \gls{MCMCFAAPF} cannot often be used in realistic scenarios because it requires sampling from $p_\theta(x_t|x_{t-1}, y_t)$ and evaluating $x_{t-1} \mapsto p_\theta(y_t|x_{t-1})$ in order to implement the \gls{FAAPF} in \eqref{eq:mcmc_fa-apf_transition_density}. To circumvent this problem, we can define a special case of the \gls{MCMCAPF} algorithm  which requires neither sampling from $p_\theta(x_t|x_{t-1}, y_t)$ nor evaluating $x_{t-1} \mapsto p_\theta(y_t|x_{t-1})$. This algorithm, obtained by setting $\tilde{p}_\theta(y|x) \equiv 1$, will be called \emph{\gls{MCMCPF}} as it represents an analogue of the (bootstrap) \gls{PF}. At time~$1$, the \gls{MCMCPF} uses the \gls{MCMC} kernels $\overline{R}_{\theta,1}$ which are invariant w.r.t.\ $\bar{\rho}_{\theta,1}(x_1) \coloneqq \mu_\theta(x_1)$. At time~$t$, $t > 1$, the \gls{MCMCPF} uses the kernels $\overline{R}_{\theta,t}(\,\cdot\,|\,\cdot\,;\mathbf{x}_{t-1})$ which are invariant w.r.t.\ 
\begin{align}
  \bar{\rho}_{\theta,t}(x_t,a_{t-1}|\mathbf{x}_{t-1}) \coloneqq \frac{g_\theta(y_{t-1}|x_{t-1}^{a_{t-1}})}{\sum_{i=1}^N g_\theta(y_{t-1}|x_{t-1}^i)} f_\theta(x_t|x_{t-1}^{a_{t-1}}). \label{eq:mcmc_fa-apf_transition_density}
\end{align}
The \gls{PMMH}-type variant of the \gls{MCMCPF} may be useful if the \gls{PMMH}-type variant of the \gls{MCMCFAAPF} cannot be implemented.

\subsection{Gibbs samplers}

Given the extended target construction of the \gls{MCMCAPF} algorithm, it is straightforward to implement \gls{PG} algorithms \gls{BS} (or similarly with \gls{AS} -- see Section~\ref{subsec:generalised_PGS}) which target it. 

However, Gibbs samplers based around the (conditional) \gls{MCMCPF} do not appear useful as they might be expected to perform less well than the Gibbs sampler based around the \gls{MCMCFAAPF} and are no more easy to implement: in contrast to the \gls{PMMH}-type algorithms, the Gibbs sampler based around the (conditional) \gls{MCMCFAAPF} does \emph{not} generally require sampling from $p_\theta(x_t|x_{t-1}, y_t)$ and it only requires evaluation of the unnormalised density $p_\theta(y_t|x_t)f_\theta(x_t|x_{t-1})$ in the transition density of the \gls{FAAPF} in \eqref{eq:mcmc_fa-apf_transition_density}.

\section{General particle Markov chain Monte Carlo methods}
\label{sec:general_pmcmc}

In this section, we describe a slight generalisation of \gls{PMCMC} methods which admits both the standard \gls{PMCMC} methods from Section~\ref{sec:pmcmc} as well as the alternative \gls{EHMM} methods from Section~\ref{sec:embeddedHMMnewversion} as special cases. In addition, we derive both the \glsdesc{BS} and \glsdesc{AS} recursions for this algorithm. We note that this section is necessarily slightly more abstract than the previous sections. As the details developed below are not required for understanding the remainder of this work, this section may be skipped on a first reading.

\subsection{Extended target distribution}
We define $\mathbf{z}_1 \coloneqq \mathbf{x}_1$ and $\mathbf{z}_t \coloneqq (\mathbf{x}_t, \mathbf{a}_{t-1})$. For notational brevity, also define $\smash{\mathbf{z}_1^{-i} \coloneqq \mathbf{z}_1 \setminus x_1^i}$, $\smash{\mathbf{z}_t^{-i} \coloneqq \mathbf{z}_t \setminus (x_t^i, a_{t-1}^i)}$ as well as $\smash{\mathbf{z}_{1:t}^{-b_{1:t}} = (\mathbf{z}_{1}^{-b_1}, \dotsc, \mathbf{z}_{t}^{-b_t})}$. We note that further auxiliary variables could be included in $\mathbf{z}_t$ without changing anything in the construction developed below. The law of a general \gls{PF} is given by
\begin{align}
 \Psi_\theta(\mathbf{z}_{1:T}) \coloneqq \psi_{\theta,1}(\mathbf{z}_1) \prod_{t=2}^T \psi_{\theta,t}(\mathbf{z}_t|\mathbf{z}_{1:t-1}).
\end{align}
With this notation, general \gls{PMCMC} methods target the following extended distribution:
\begin{align}
 \tilde{\pi}(\theta, \mathbf{z}_{1:T}, b_T) \coloneqq \frac{1}{N^T} \times \underbrace{\pi(\theta, x_{1:T}^{b_{1:T}})}_{\text{\footnotesize{target}}} \times \underbrace{\phi_\theta(\mathbf{z}_{1:T}^{-b_{1:T}} | x_{1:T}^{b_{1:T}}, b_{1:T})}_{\text{\parbox{2.3cm}{\centering\footnotesize{law of conditional general \gls{PF}}}}}, \label{eq:general_pmcmb_target_distribution}
\end{align}
where the law of the conditional general \gls{PF} is given by
\begin{align}
 \phi_\theta(\mathbf{z}_{1:T}^{-b_{1:T}} | x_{1:T}^{b_{1:T}}, b_{1:T}) \coloneqq \psi_{\theta,1}^{-b_1}(\mathbf{z}_1^{-b_1}) \prod_{t=2}^T  \psi_{\theta,t}^{-b_t}(\mathbf{z}_t^{-b_t}|\mathbf{z}_{1:t-1}, x_t^{b_t}),
\end{align}
with
\begin{align}
 \psi_{\theta,t}^{-i}(\mathbf{z}_t^{-i}|\mathbf{z}_{1:t-1}, x_t^i)
 & \coloneqq \frac{\psi_{\theta,t}(\mathbf{z}_t|\mathbf{z}_{1:t-1})}{\psi_{\theta,t}^i(x_t^i, a_{t-1}^i|\mathbf{z}_{1:t-1}, x_t^i)}.
\end{align}
Here, $\psi_{\theta,t}^i(\,\cdot\,|\mathbf{z}_{1:t-1})$ denotes the marginal distribution of the $i$th components of $\mathbf{x}_t$ and $\mathbf{a}_{t-1}$ under the distribution $\psi_{\theta,t}(\,\cdot\,|\mathbf{z}_{1:t-1})$. 
Finally, for any $t \in \{1,\dotsc,T\}$, we define the following unnormalised weight
 \begin{align}
 \tilde{v}_{\theta, t}^{b_t} \coloneqq \frac{1}{N^t} \frac{\gamma_{\theta,t}(x_{1:t}^{b_{1:t}})}{\psi_{\theta,1}^{b_1}(x_1^{b_1}) \prod_{n=2}^t \psi_{\theta,n}^{b_n}(x_n^{b_n}, a_{n-1}^{b_n}|\mathbf{z}_{1:n-1})},
\end{align}
where $b_{1:t-1}$ on the r.h.s.\ are to be interpreted as functions of $b_t$ and the ancestry variables via the usual recursion $\smash{b_t = a_t^{b_{t+1}}}$. Here, $\gamma_{\theta,t}(x_{1:t})$ is the unnormalised density targeted at the $t$th step of the general \gls{PF} -- for all the algorithms discussed in this work, we will state these densities explicitly in Appendix~\ref{subsec:special_cases}; in particular,
\begin{equation}
 \gamma_{\theta,T}(x_{1:T}) = p_\theta(x_{1:T}, y_{1:T}).
\end{equation}

We make the following minimal assumption to ensure the validity of the (general) \gls{PMCMC} algorithms.

\begin{assumption}[absolute continuity] \label{as:absolute_continuity}
 For any $t \in \{1,\dotsc,T\}$, any $i \in \{1,\dotsc,N\}$ and any $\mathbf{z}_{1:t-1}$, the support of $(x_t, b_{t-1}) \mapsto \psi_{\theta,t}^i(x_t, b_{t-1}|\mathbf{z}_{1:t-1})$ includes the support of $(x_t, b_{t-1}) \mapsto \gamma_{\theta,t}(x_{1:t-1}^{b_{1:t-1}}, x_{t})$. 
\end{assumption}

We also make the following assumption which requires that all marginals of the conditional distributions $\psi_{\theta,t}(\,\cdot\,|\mathbf{z}_{1:t-1})$ are identical.

\begin{assumption}[identical marginals] \label{as:exchangeability}
 For any $(i,j) \in \{1,\dotsc, N\}^2$ and any $t \in \{1,\dotsc,T\}$, $\psi_{\theta,t}^i = \psi_{\theta,t}^j$.
\end{assumption}

\begin{remark}\label{rem:identical_marginals_assumption}
 Assumption~\ref{as:exchangeability} can be easily dropped in favour of selecting a suitable (non-uniform) distribution for the particle indices $b_{1:T}$ in \eqref{eq:general_pmcmb_target_distribution}. Indeed, more elaborate constructions could be used to justify resampling schemes which, unlike multinomial resampling, are not exchangeable in the sense of \citet[Assumption~2]{AndrieuDoucetHolenstein2010} (unless one permutes the particle indices uniformly at random at the end of each step as mentioned in \citet{AndrieuDoucetHolenstein2010}). Similarly, such more general constructions would allow us to view the use of more sophisticated \glspl{PF}, such as the \emph{discrete particle filter} of \citet{Fearnhead1998}, with \gls{PMCMC} schemes as special cases of this framework as shown in \citet[Section~2.3.4]{Finke2015}.
\end{remark}


 In Examples~\ref{ex:antithetic} and \ref{ex:sqmc}, we show how \glspl{APF} with \emph{antithetic variables} \citep{BizjajevaOlsson2008} and (randomised) \emph{\gls{SQMC} methods} \citep{GerberChopin2015} can be considered as special cases of the framework described in this section even though these methods cannot easily be viewed as conventional \glspl{PF} because the particles are not sampled conditionally independently at each step. 
 
 \begin{example}[\glspl{APF} with antithetic variables] \label{ex:antithetic}
  The \glspl{APF} with antithetic variables from \citet{BizjajevaOlsson2008} aim to improve the performance of \glspl{APF} by introducing negative correlation into the particle population. To that end, the $N$ particles are divided into $M$ groups of $K$ particles; the particles in each group then share the same ancestor index and given the ancestor particle, they are sampled in such a way that they are negatively correlated.
  
  Assume that there exists $K, M \in \mathbb{N}$ such that $N = K M$ and for $\tilde{x}_t \coloneqq (\tilde{x}_t^1, \dotsc, \tilde{x}_t^K)  \in \mathcal{X}^K$ let $\tilde{q}_{\theta,t}(\tilde{x}_t| x_{t-1})$ denote some joint proposal kernel for $K$ particles such that if $(\tilde{x}_t^1, \dotsc, \tilde{x}_t^K) \sim \tilde{q}_{\theta,t}(\,\cdot\,| x_{t-1})$ then
  \begin{enumerate*}
   \item $\tilde{x}_t^1, \dotsc, \tilde{x}_t^K$ are (pairwise) negatively correlated,
   \item marginally, $\smash{\tilde{x}_t^k \sim q_{\theta,t}(\,\cdot\,| x_{t-1})}$ for all $k \in \{1,\dotsc,K\}$.
  \end{enumerate*}
  
  Given $\mathbf{z}_{1:t-1}$, the \gls{APF} with antithetic variables generates $\mathbf{z}_t = (\mathbf{a}_{t-1}, \mathbf{x}_t)$ as follows (we use the convention that any action prescribed for some $m$ is to be performed for all $m \in \{1,\dotsc,M\}$).
  \begin{enumerate}
   \item \label{ex:antithetic:step:1} Set $\smash{a_{t-1}^{(m-1)K+1} = i}$ w.p.\ proportional to $v_{\theta, t-1}^i$.
   \item Set $\smash{a_{t-1}^{(m-1)K+k} \coloneqq a_{t-1}^{(m-1)K+1}}$ for all $k \in \{2,\dotsc,K\}$.
   \item \label{ex:antithetic:step:2} Sample $\smash{\bigl(x_t^{(m-1)K+k}\bigr)_{k \in \{1,\dotsc,K\}} \sim \tilde{q}_{\theta,t}\bigl(\,\cdot\,\big| x_{t-1}^{a_{t-1}^{(m-1)K+1}}\bigr)}$.
   \item \label{ex:antithetic:step:3} Permute the particle indices on $\mathbf{z}_t^1, \dotsc, \mathbf{z}_t^N$ uniformly at random.
  \end{enumerate}
 \end{example}

\begin{example}[sequential quasi Monte Carlo]\label{ex:sqmc}
 Let $\mathcal{X} = \mathbb{R}^d$. Randomised \gls{SQMC} algorithms are general \glspl{PF} which stratify sampling of the ancestor indices and particles $\mathbf{z}_t = (\mathbf{a}_{t-1}, \mathbf{x}_t)$ by computing them as a deterministic transformation of a set of randomised quasi Monte Carlo points $\mathbf{u}_t \coloneqq (u_t^1, \dotsc, u_t^N) \in [0,1)^{(d+1)N}$. By construction, 
 \begin{enumerate*}
  \item \label{ex:sqmc:property:1} the set $\mathbf{u}_t = (u_t^1, \dotsc, u_t^N)$ has a low discrepancy, 
  \item \label{ex:sqmc:property:2} for each $i \in \{1,\dotsc,N\}$, $u_t^i$ is (marginally) uniformly distributed on the $(d+1)$-dimensional hypercube.
 \end{enumerate*}
 
 Write $u_t^i = (\tilde{u}_t^i, \tilde{v}_t^i)$ with $\tilde{u}_t^i \in [0,1)$ and $\tilde{v}_t^i \in [0,1)^{d}$. Given $\mathbf{z}_{1:t-1}$, the algorithm \citep[Algorithm~3]{GerberChopin2015} transforms $\mathbf{u}_t \to \mathbf{z}_t = (\mathbf{a}_{t-1}, \mathbf{x}_t)$ as follows (using the convention that any action mentioned for some $i$ is to be performed for all $i \in \{1,\dotsc,N\}$).
 \begin{enumerate}
  \item \label{ex:sqmc:step:1} Find a suitable permutation $\sigma_{t-1}\colon \{1,\dotsc, N\} \to \{1,\dotsc,N\}$ such that $x_{t-1}^{\sigma_{t-1}(1)} \leq \dotsc \leq x_{t-1}^{\sigma_{t-1}(N)}$, if $d=1$; if $d > 1$, the permutation $\sigma_{t-1}$ is obtained by mapping the particles to the hypercube $[0,1)^{d}$ and projecting them onto $[0,1)$ using the pseudo-inverse of the Hilbert space-filling curve. These projections are then ordered as for $d=1$ (see \citet{GerberChopin2015} for details).
  
  \item \label{ex:sqmc:step:2} Set $\smash{a^i \coloneqq F^{-1}(\tilde{u}_t^i)}$, where $F^{-1}$ denotes the generalised inverse of the \gls{CDF} $F\colon \{1,\dotsc,N\} \to [0,1]$, defined by $\smash{F(i) \coloneqq \sum_{j=1}^i v_{\theta,t-1}^{\sigma_{t-1}(j)} / \sum_{j=1}^N v_{\theta,t-1}^j}$. 
  
  \item \label{ex:sqmc:step:3} Set $\smash{a_{t-1}^i \coloneqq \sigma_{t-1}(a^i)}$ and $\smash{x_t^i \coloneqq \varGamma_{\theta,t}(x_{t-1}^{a_{t-1}^i}, \tilde{v}_t^i)}$. Here, if $d=1$, the function $\varGamma_{\theta,t}(x_{t-1}, \,\cdot\,)$ is the (generalised) inverse of the \gls{CDF} associated with $q_{\theta,t}(\,\cdot\,|x_{t-1})$; if $d > 1$, this can be generalised via the Rosenblatt transform.
  
  \item \label{ex:sqmc:step:4} Permute the particle indices on $\mathbf{z}_t^1, \dotsc, \mathbf{z}_t^N$ uniformly at random.
 \end{enumerate}
\end{example}

While the joint kernel $\psi_{\theta,t}(\mathbf{z}_t|\mathbf{z}_{1:t-1})$ is potentially intractable in both examples, the random permutation of the particle indices (i.e.\ Step~\ref{ex:antithetic:step:3} in Example~\ref{ex:antithetic} and also Step~\ref{ex:sqmc:step:4} in Example~\ref{ex:sqmc}) ensures that Assumption~\ref{as:exchangeability} is satisfied. Indeed, it can be easily verified that in both examples, for any $(i,j) \in \{1,\dotsc,N\}^2$,
 \begin{align}
  \psi_{\theta,t}^i(x_t,a_{t-1}|\mathbf{z}_{1:t-1}) 
   = \rho_{\theta,t}^{\mathbf{q}_\theta}(x_t, a_{t-1}|\mathbf{x}_{t-2:t-1},  \mathbf{a}_{t-2})
   = \psi_{\theta,t}^j(x_t, a_{t-1}|\mathbf{z}_{1:t-1}).
 \end{align}
 As pointed out in Remark~\ref{rem:identical_marginals_assumption}, Assumption~\ref{as:exchangeability} is not actually necessary and can be easily dropped in favour of a slightly more general construction of the extended target distribution which is implicitly employed by \citet{BizjajevaOlsson2008, GerberChopin2015} (who therefore do not require the random permutation of the particle indices).

\subsection{General particle marginal Metropolis--Hastings}

In this section, we use the general \gls{PMCMC} framework to derive a general \gls{PMMH} algorithm. All \gls{PMMH} algorithms and \gls{MH} versions of the alternative \gls{EHMM} methods can then be seen as special cases of this general scheme as shown in  Appendix~\ref{subsec:special_cases}.
As with the standard \gls{PMMH}, we may use an \gls{MH} algorithm to target the extended distribution $\tilde{\pi}(\theta, \mathbf{z}_{1:T}, b_T)$ using a proposal of the form
\begin{equation}
  q(\theta,{\theta'}) \times \underbrace{\Psi_{{\theta'}}(\mathbf{z}_{1:T})}_{\text{\footnotesize{\parbox{1.5cm}{\centering law of general \gls{PF}}}}} \times \underbrace{q_{\theta'}(b_T|\mathbf{z}_{1:T})}_{\text{\footnotesize{path selection}}}, \label{eq:proposalGeneralisedPF}
\end{equation}
where we have defined the selection probability
\begin{align}
 q_\theta(b_T|\mathbf{z}_{1:T}) \coloneqq \frac{\tilde{v}_{\theta, T}^{b_T}}{\sum_{i=1}^N \tilde{v}_{\theta, T}^i}.
\end{align}
Define the usual unbiased estimate of the marginal likelihood
\begin{equation}
 \hat{p}_\theta(y_{1:T}) \coloneqq \sum_{i=1}^N \tilde{v}_{\theta, T}^i.
\end{equation}
Then we obtain the following general \gls{PMMH} algorithm (Algorithm~\ref{alg:general_pmmh}) the validity of which can be established by checking that indeed,
\begin{align}
  \frac{\tilde{\pi}(\theta, \mathbf{z}_{1:T}, b_T)}{\Psi_\theta(\mathbf{z}_{1:T}) q_\theta(b_T|\mathbf{z}_{1:T})} 
 = p(\theta|y_{1:T}) \frac{\hat{p}_\theta(y_{1:T})}{p_\theta(y_{1:T})}.
\end{align}

\noindent\parbox{\textwidth}{
\begin{flushleft}
\begin{framedAlgorithm}[general \gls{PMMH} algorithm] \label{alg:general_pmmh} Given $(\theta, \mathbf{z}_{1:T}, b_T) \sim \tilde{\pi}(\theta, \mathbf{z}_{1:T}, b_T)$ with associated likelihood estimate $\hat{p}_{\theta}(y_{1:T})$.
 \begin{enumerate}
  \item Propose $\theta' \sim q(\theta, \theta')$, $\mathbf{z}_{1:T}' \sim \Psi_{\theta'}(\mathbf{z}_{1:T}')$ and $b_T' \sim q_{\theta'}(b_T'|\mathbf{z}_{1:T}')$.
  \item Compute likelihood estimate $\hat{p}_{\theta'}(y_{1:T})$ based on $\mathbf{z}_{1:T}'$.
  \item Set $(\theta, \mathbf{z}_{1:T}, b_T) \leftarrow (\theta', \mathbf{z}_{1:T}', b_T')$ w.p.\
  $
  1 \wedge 
  \dfrac{\smash{\hat{p}_{{\theta'}}(y_{1:T})p({\theta'})}}{\hat{p}_{\theta}(y_{1:T})p(\theta)}
  \dfrac{\smash{q({\theta'},\theta)}}{q(\theta,{\theta'})}
  . \label{eq:generalisedPMMHratio}
  $
 \end{enumerate}
\end{framedAlgorithm}
\end{flushleft}
}

\subsection{General particle Gibbs samplers}
\label{subsec:generalised_PGS}
\glsreset{AS}
\glsreset{BS}

In this section, we use the general \gls{PMCMC} framework to derive a general \gls{PG} sampler. We also derive \gls{BS} \citep{Whiteley2010} and \gls{AS} \citep{LindstenJordanSchon2014} recursions and prove that they leave the target distribution of interest invariant. As before, all \gls{PG} samplers and Gibbs versions of the alternative \gls{EHMM} method can then be seen as special cases of this general scheme as shown in Appendix~\ref{subsec:special_cases}. Set
\begin{equation}
 \gamma_\theta(x_{t+1:T}|x_{1:t}) \coloneqq \frac{\gamma_{\theta,T}(x_{1:T})}{\gamma_{\theta,t}(x_{1:t})}.
\end{equation}
We are then ready to state both (general) \gls{PG} samplers. For the remainder of this section, we let $\tilde{x}_{1:t}^i$ denote the $i$th particle lineage at time~$t$, i.e.\ $\smash{\tilde{x}_{1:t}^i = x_{1:t}^{i_{1:t}}}$, where $i_t = i$ and $\smash{i_n = a_n^{i_{n+1}}}$, for $n = t-1, \dotsc, 1$.

\noindent\parbox{\textwidth}{
\begin{flushleft}
\begin{framedAlgorithm}[general \gls{PG} sampler with \gls{BS}] \label{alg:general_pg_bs} Given $(\theta, x_{1:T}) \sim \pi$, obtain $(\theta', x_{1:T}') \sim \pi$ as follows.
 \begin{enumerate}
  \item Sample $\theta'$ via some $\pi(\,\cdot\,|x_{1:T})$-invariant \gls{MCMC} kernel.
  \item For $t = 1,\dotsc,T$, perform the following steps.
  \begin{enumerate}
   \item If $t=1$, sample $b_1$ uniformly on $\{1,\dotsc,N\}$, set $x_1^{b_1} \coloneqq x_1$ and sample $\mathbf{z}_1^{-b_1} \sim \psi_{\theta',1}^{-b_t}(\mathbf{z}_1^{-b_1}|x_1^{b_1})$.
   \item If $t>1$, sample $b_t$ uniformly on $\{1,\dotsc,N\}$, set $x_t^{b_t} \coloneqq x_t$, $a_{t-1}^{b_t} \coloneqq b_{t-1}$ and sample $\mathbf{z}_t^{-b_t} \sim \psi_{{\theta'},t}^{-b_t}(\mathbf{z}_t^{-b_t}|\mathbf{z}_{1:t-1}, x_t^{b_t})$.
  \end{enumerate}
  \item Sample $b_T \sim q_{\theta'}(b_T|\mathbf{z}_{1:T})$ and for $t = T-1, \dotsc, 1$, set $b_t=i$ w.p.\ proportional to $\tilde{v}_{{\theta'}, t}^{i}\gamma_{\theta'}(x_{t+1:T}^{b_{t+1:T}}|\tilde{x}_{1:t}^i)$.
  \item Set $x_{1:T}' \coloneqq x_{1:T}^{b_{1:T}}$.
 \end{enumerate}
\end{framedAlgorithm}
\end{flushleft}
}

\noindent\parbox{\textwidth}{
\begin{flushleft}
\begin{framedAlgorithm}[general \gls{PG} sampler with \gls{AS}] \label{alg:general_pg_as} Given $(\theta, x_{1:T}) \sim \pi$, obtain $(\theta', x_{1:T}') \sim \pi$ as follows.
 \begin{enumerate}
  \item Sample $\theta'$ via some $\pi(\,\cdot\,|x_{1:T})$-invariant \gls{MCMC} kernel.
  \item For $t = 1,\dotsc,T$, perform the following steps.
  \begin{enumerate}
   \item If $t=1$, sample $b_1$ uniformly on $\{1,\dotsc,N\}$, set $x_1^{b_1} \coloneqq x_1$ and sample $\mathbf{z}_1^{-b_1} \sim \psi_{\theta',1}^{-b_t}(\mathbf{z}_1^{-b_1}|x_1^{b_1})$.
   \item If $t>1$, sample $b_t$ uniformly on $\{1,\dotsc,N\}$, set $x_t^{b_t} \coloneqq x_t$, set $a_{t-1}^{b_t} = i$ w.p.\ proportional to $\tilde{v}_{{\theta'}, t-1}^{i}\gamma_{\theta'}(x_{t:T}^{b_{t:T}}|\tilde{x}_{1:t-1}^{i})$ and sample $\mathbf{z}_t^{-b_t} \sim \psi_{{\theta'},t}^{-b_t}(\mathbf{z}_t^{-b_t}|\mathbf{z}_{1:t-1}, x_t^{b_t})$.
  \end{enumerate}
  \item Sample $b_T \sim q_{\theta'}(b_T|\mathbf{z}_{1:T})$ and for $t = T-1, \dotsc, 1$, set $b_t \coloneqq a_t^{b_{t+1}}$.
  \item Set $x_{1:T}' \coloneqq x_{1:T}^{b_{1:T}}$.
 \end{enumerate}
\end{framedAlgorithm}
\end{flushleft}
}

As in previous sections, the \gls{BS} recursion in Algorithm~\ref{alg:general_pg_bs} may be justified via appropriate partially-collapsed Gibbs sampler arguments by noting that
\begin{align}
 \tilde{\pi}(b_t|\theta, \mathbf{z}_{1:t}, x_{t+1:T}^{b_{t+1:T}})
 & \propto \tilde{v}_{{\theta}, t}^{b_t}\gamma_{\theta}(x_{t+1:T}^{b_{t+1:T}}|\tilde{x}_{1:t}^{b_t}).
\end{align}
The \gls{AS} steps in Algorithm~\ref{alg:general_pg_as} follows similarly since $a_t^{b_{t+1}} = b_t$, by construction. 

Alternatively -- without invoking partially-collapsed Gibbs sampler arguments -- the validity of \gls{BS} can be established by even further extending the space to include the new particle indices generated via \gls{BS}. As shown in \citet[Chapter~3.4.3]{Finke2015}, this construction also proves a particular duality of \gls{BS} and \gls{AS}.

\section{Empirical study}
\label{sec:simulations}

In this section, we empirically compare the performance of some of the algorithms described in this work on a $d$-dimensional linear-Gaussian state-space model.

\subsection{Model}
 The model considered throughout this section is given by
\begin{align}
 \mu_\theta(x_1) & = \mathrm{Normal}(x_1; m_0, C_0),\\
 f_\theta(x_t|x_{t-1}) & = \mathrm{Normal}(x_t; A x_{t-1}, \sigma^2 \mathbf{I}_d), \quad \text{for $t > 1$,}\\
 g_\theta(y_t|x_t) & = \mathrm{Normal}(y_t; x_t, \tau^2 \mathbf{I}_d), \quad \text{for $t \geq 1$,}
\end{align}
where $x_t, y_t \in \mathbb{R}^d$, $\sigma, \tau > 0$,  $\mathbf{I}_d$ denotes the $(d,d)$-dimensional identity matrix and $A$ is the $(d,d)$-dimensional symmetric banded matrix with upper and lower bandwidth $1$, with entries $a_0 \in \mathbb{R}$ on the main diagonal, and with entries $a_1 \in \mathbb{R}$ on the remaining bands, i.e.\ 
\begin{equation}
 A = 
 \begin{bmatrix}
  a_0 & a_1 & 0 & \dotsc & 0\\
  a_1 & a_0 & a_1 & \ddots & \vdots\\
  0 & a_1 & \ddots & \ddots & 0\\
  \vdots & \ddots & \ddots & a_0 & a_1\\
  0 & \dotsc & 0 &  a_1 & a_0
 \end{bmatrix}.
\end{equation}
For simplicity, we assume that the initial mean $m_0 \coloneqq \mathbf{0}_d \in \mathbb{R}^d$ (where $\mathbf{0}_d$ denotes a vector of zeros of length~$d$) and the initial $(d,d)$-dimensional covariance matrix $C_0 = \mathbf{I}_d$ are known. Thus, the task is to approximate the posterior distribution of the remaining parameters $\theta \coloneqq (a_0, a_1, \sigma, \tau)$. The true values of these parameters, i.e.\ the values used for simulating the data are $(0.5, 0.2, 1, 1)$. As prior distributions, we take uniform distributions on $(-1,1)$ for $a_0$ and $a_1$ and inverse-gamma distributions on $\sigma$ and $\tau$ each with shape parameter $1$ and scale parameter $0.5$. All parameters are assumed to be independent a priori. In all algorithms, we propose new values ${\theta'}$ for $\theta$ via a simple Gaussian random-walk kernel, i.e.\ we use $q(\theta, {\theta'}) \coloneqq \mathrm{Normal}({\theta'};\theta, (100 d_\theta d T)^{-1} \mathbf{I}_{d_\theta})$, where $d_\theta$ is the dimension of the parameter vector $\theta$, i.e.\ $d_\theta = 4$.

\subsection{Algorithms}
\label{subsec:algorithms}

In this subsection, we detail the specific algorithms whose empirical performance we compare in our simulation study.

\begin{description}[leftmargin=0pt, labelindent=0pt]
  \item[Standard \gls{PMCMC}.] We implement the (bootstrap) \gls{PF} and the \gls{FAAPF} using multinomial resampling at every step. Though we note that more sophisticated resampling schemes, e.g.\ adaptive systematic resampling, could easily be employed. As described above, we can implement both \gls{MH} algorithms (i.e.\ the \gls{PMMH}) and Gibbs samplers based around these standard \glspl{PF}. For the latter, we make use of \gls{AS} in the conditional \glspl{PF}.

  \item[Original \gls{EHMM}.] We implement the algorithms with $\rho_{\theta,t}(x) = \mathrm{Normal}(x; \mu, \varSigma)$, where $\mu$ and $\varSigma$ represent the mean and covariance matrix associated with the stationary distribution of the latent Markov chain $(X_t)_{t \in \mathbb{N}}$. We compare two different options for constructing the kernels $R_{\theta,t}$ which leave this distribution invariant.
  \begin{enumerate}[label=(\Roman*), ref=\Roman*]
    \item \label{as:original_EHMM_kernel_a} The kernel $R_{\theta,t}$ generates \gls{IID} samples from its invariant distribution, i.e.\ $\smash{R_{\theta,t}(x_t'|x_t) = \rho_{\theta,t}(x_t')}$.
    \item \label{as:original_EHMM_kernel_b} The kernel $R_{\theta,t}$ is a standard \gls{MH} kernel which proposes a value $x_t^\star$ using the Gaussian random-walk proposal $\smash{\mathrm{Normal}(x_t^\star; x_t, d^{-1}\mathbf{I}_d)}$.
  \end{enumerate}

  \item[Alternative \gls{EHMM}.] We compare four different versions of the \gls{MCMCPF} and \gls{MCMCFAAPF} methods outlined above. Again, we implement both \gls{MH} algorithms and Gibbs samplers (with \gls{AS}) based around these methods. Below, we describe the specific versions which we are comparing. The kernels $\overline{R}_{\theta, t}(\,\cdot\,|\,\cdot\,;\mathbf{x}_{t-1})$ employed in the \gls{MCMCPF} and the kernels $R_{\theta, t}(\,\cdot\,|\,\cdot\,;\mathbf{x}_{t-1})$ employed in the \gls{MCMCFAAPF} are all taken to be \gls{MH} kernels which, given $(x_t, a_{t-1})$, propose a new value $(x_t^\star, a_{t-1}^\star)$ using a proposal of the following form 
  \begin{equation}
  \frac{v_{\theta,t-1}^{a_{t-1}^\star}}{\sum_{i=1}^N v_{\theta,t-1}^{i}} s_{\theta,t}(x_t^\star|x_t;\mathbf{x}_{t-1}, a_{t-1}^\star).
  \end{equation}
  We compare two different approaches for generating a new value for the particle, $x_t^\star$.
  \begin{enumerate}[label=(\Roman*), ref=\Roman*]
  \item \label{as:particle_proposal_a} The first proposal uses a simple Gaussian random-walk kernel, i.e.\
  \begin{equation}
    s_{\theta,t}(x_t^\star|x_t;\mathbf{x}_{t-1}, a_{t-1}^\star) = \mathrm{Normal}(x_t^\star; x_t, d^{-1}\mathbf{I}_d), \label{eq:rw_proposal}
  \end{equation}
  where the scaling of the covariance matrix is motivated by existing results on optimal scaling for such random-walk proposal kernels \citep{GelmanRobertsGilks1996,RobertsGelmanGilks1997}.
%
  \item \label{as:particle_proposal_b} The second proposal uses the \emph{autoregressive} proposal employed by \cite{ShestopaloffNeal2016}, i.e.\ 
  \begin{equation}
    s_{\theta,t}(x_t^\star|x_t;\mathbf{x}_{t-1}, a_{t-1}^\star) = \mathrm{Normal}\bigl(x_t^\star;\mu + \sqrt{1-\varepsilon^2} (x_t - \mu), \varepsilon^2\varSigma\bigr), \label{eq:ar_proposal}
  \end{equation}
  where $\mu$ and $\varSigma$ denote the mean and covariance matrix of $f_\theta(x_t|x_{t-1}) = \mathrm{Normal}(x_t; \mu, \varSigma)$, i.e.\ $\varSigma=\sigma^2\mathbf{I}$ and $\mu=A x_{t-1}$. To scale the covariance matrix of this proposal with the dimension $d$, we set $\varepsilon \coloneqq \sqrt{d^{-1}}$.
  
  \end{enumerate}

  \item[Idealised.] We also implement the algorithms which the above-mentioned algorithms seek to mimic. The idealised Gibbs sampler, is a (Metropolis-within-)Gibbs algorithm which updates the latent states $x_{1:T}$ as one block by sampling them from their full conditional posterior distribution. The idealised marginal \gls{MH} algorithm analytically evaluates the marginal likelihood $p_\theta(y_{1:T})$ via the Kalman filter.
\end{description}

\subsection{Results for general PMMH algorithms}
In this subsection, we empirically compare the performance of various \gls{PMMH} type samplers. First, we fix $\theta$ in order to assess the variability of the estimates of the marginal likelihood, $\hat{p}_\theta(y_{1:T})$, which is a key ingredient in (general) \gls{PMMH} algorithms. Then, we perform inference about $\theta$.

Recall that in order to implement the \gls{MH} version of the \gls{MCMCFAAPF}, we need to sample at least one particle from $p_\theta(x_t|x_{t-1}, y_t)$ at each time~$t$ and we need to be able to evaluate the function $x_{t-1} \mapsto p_\theta(y_t|x_{t-1})$. In other words, whenever we can implement this algorithm we can also implement a standard \gls{PMMH} algorithm based around the \gls{FAAPF}.

\begin{figure}[H]
  \noindent{}
  \centering
  \begin{subfigure}[t]{0.188\linewidth}
    \centering
      \includegraphics[trim = 4cm 0cm 2cm 0cm]{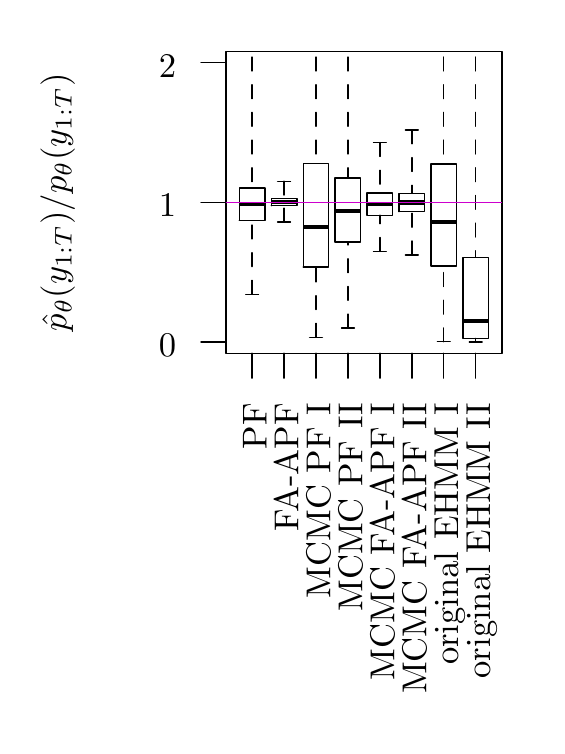}
    \caption{$d=2$.}\label{fig:likelihood_estimates_in_dimension_2}		
  \end{subfigure}
  \begin{subfigure}[t]{0.188\linewidth}
    \centering
      \includegraphics[trim = 4cm 0cm 2cm 0cm]{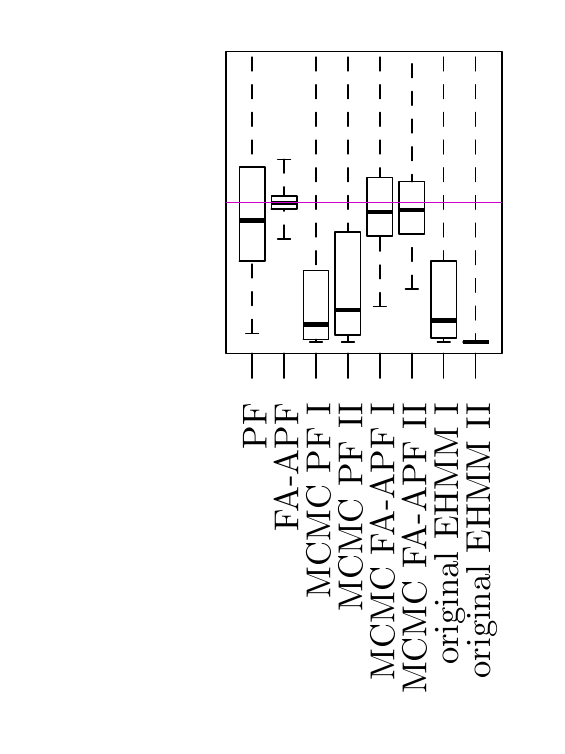}
    \caption{$d=5$.}\label{fig:likelihood_estimates_in_dimension_5}
  \end{subfigure}
  \begin{subfigure}[t]{0.188\linewidth}
    \centering
      \includegraphics[trim = 4cm 0cm 2cm 0cm]{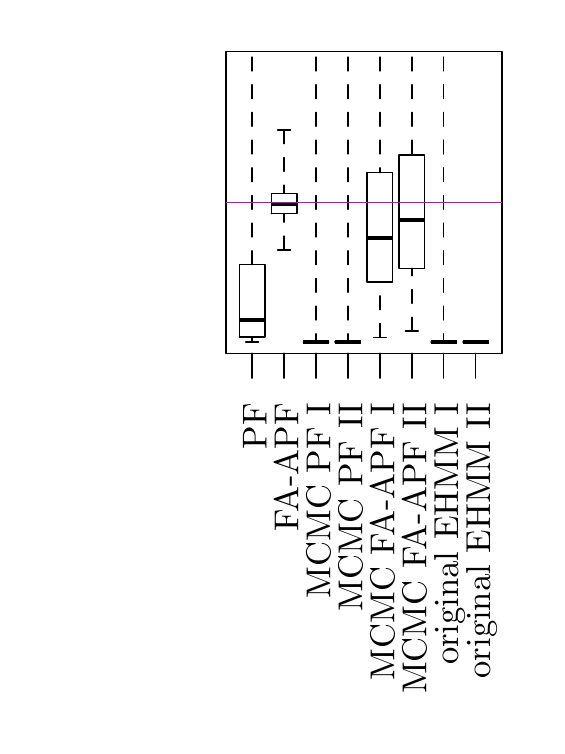}
    \caption{$d=10$.}\label{fig:likelihood_estimates_in_dimension_10}		
  \end{subfigure}
  \begin{subfigure}[t]{0.188\linewidth}
    \centering
      \includegraphics[trim = 4cm 0cm 2cm 0cm]{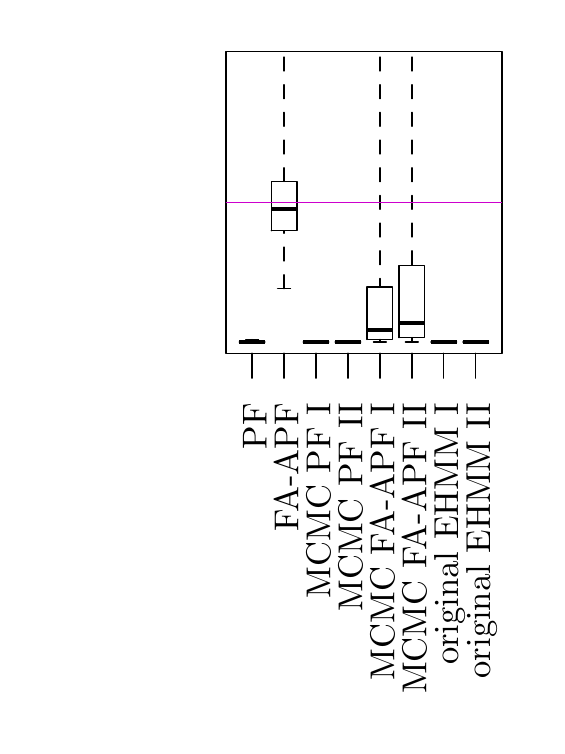}
    \caption{$d=25$.}\label{fig:likelihood_estimates_in_dimension_25}
  \end{subfigure}
  \caption{Relative estimates of the marginal likelihood $p_\theta(y_{1:T})$. Based on $1\,000$ independent runs of each algorithm (and writing $\hat{p}_\theta(y_{1:T}) = \tilde{p}_\theta(y_{1:T})$ in the case of the original \gls{EHMM} method) with each run using a different data sequence of length $T=10$ simulated from the model. The number of particles was $N=1\,000$ for the $\mathrm{O}(N)$ methods and $N=100$ for the $\mathrm{O}(N^2)$ methods.}
  \label{fig:likelihood_estimates}
\end{figure}

Figure~\ref{fig:likelihood_estimates} shows the relative estimates of the marginal likelihood obtained from the various algorithms described in this work for various model dimensions. Unsurprisingly, the \gls{PF}, resp.\ \gls{FAAPF}, provides lower variance estimates than its corresponding \gls{MCMCPF}, resp.\ \gls{MCMCFAAPF} counterparts. However, more interestingly, the \gls{MCMCFAAPF} can provide lower variance estimates than the standard \gls{PF} and could prove useful in more realistic scenarios where it is computationally very expensive to run the \gls{FAAPF}. As expected, the original \gls{EHMM} method described in Section~\ref{sec:embeddedHMM} breaks down very quickly as the dimension $d$ increases. 

\begin{figure}[H]
  \noindent{}
  \centering
  \begin{subfigure}[t]{\linewidth}
    \centering
     \includegraphics[trim = 1cm 0cm 0cm 0.25cm]{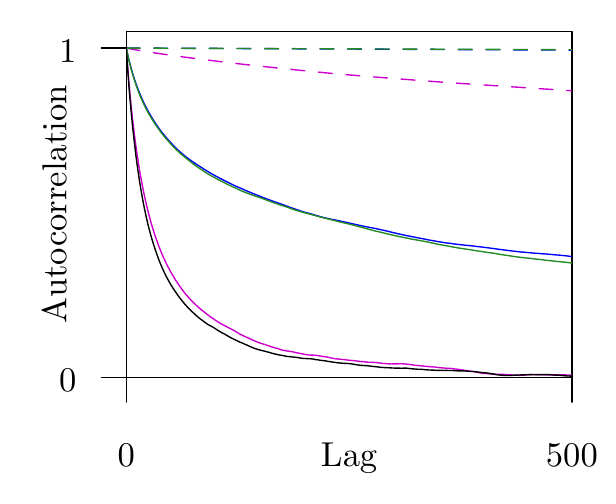}
     \includegraphics[trim = 0cm 0cm 0cm 0.25cm]{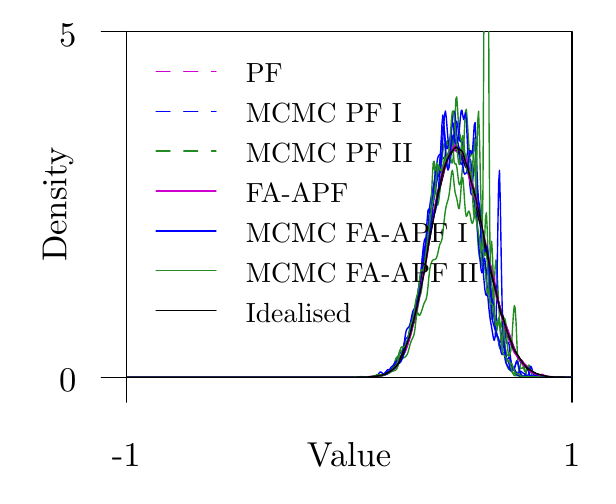}
  \end{subfigure}

  \caption{Autocorrelation (left panel) and kernel-density estimate (right panel) of the estimates of Parameter~$a_0$ for model dimension~$d=25$ and with $T=10$ observations. Obtained from $10^6$ iterations (of which the initial $10$~\% were discarded as burn-in) of standard \gls{PMMH} algorithms and \gls{MH} versions of the alternative \gls{EHMM} method using $N=1000$ particles. The autocorrelations shown on the r.h.s.\ are averages over four independent runs of each algorithm. \textit{Note:} the \gls{PMMH} algorithms based on the (bootstrap) \gls{PF} and based on the \gls{MCMCPF} failed to yield meaningful approximations of the posterior distribution and the corresponding kernel-density estimates are therefore suppressed.}
  \label{fig:pmmh_acf_parameters}
\end{figure}

The right panel of Figure~\ref{fig:pmmh_acf_parameters} shows kernel-density plots of the estimates of parameter~$a_0$ obtained from various \gls{PMMH}-type algorithms. Clearly, the \gls{PMMH}-type algorithms based around the (bootstrap) \gls{PF} or the \gls{MCMCPF} were unable to obtain sensible parameter estimates within the number of iterations that we fixed. The left panel of Figure~\ref{fig:pmmh_acf_parameters} shows the corresponding empirical autocorrelation. The results are consistent with the efficiency of the likelihood estimates illustrated in Figure~\ref{fig:likelihood_estimates}. That is, at least in this setting, the standard \gls{MH} version of the alternative \gls{EHMM} method does not outperform standard \gls{PMMH} algorithms. The estimates of the other parameters behaved similarly and the results for $(a_1, \sigma, \tau)$ are therefore omitted.

\subsection{Results for general particle Gibbs samplers}

In this subsection, we compare empirically the performance of various \gls{PG} type samplers (all using \gls{AS}). Gibbs samplers based on the original \gls{EHMM} method failed to yield meaningful estimates for the model dimensions considered in this subsection and at a similar computational cost as the other algorithms. We therefore do not show results for the original \gls{EHMM} method in the figures below.

Recall that in order to implement the conditional \gls{MCMCFAAPF}, we do not need to sample from $p_\theta(x_t|x_{t-1}, y_t)$ nor evaluate the function $x_{t-1} \mapsto p_\theta(y_t|x_{t-1})$. In other words, we can implement the conditional \gls{MCMCFAAPF} in many situations in which implementing a standard conditional \gls{FAAPF} is impossible. 

\begin{figure}[H]
  \noindent{}
  \centering
     \includegraphics[trim = 1cm 0cm 0cm 0.25cm]{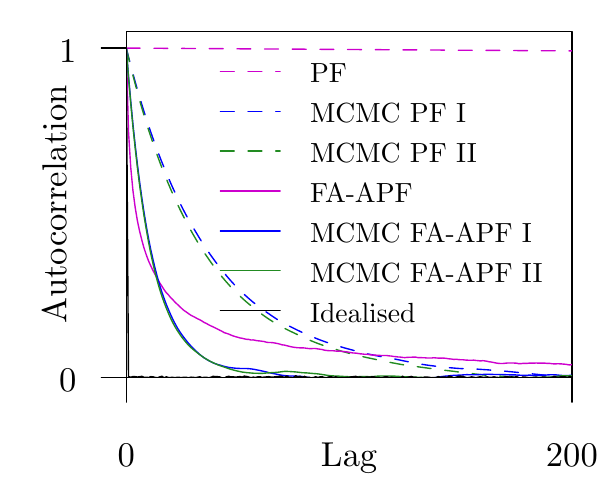}
     \includegraphics[trim = 0cm 0cm 0cm 0.25cm]{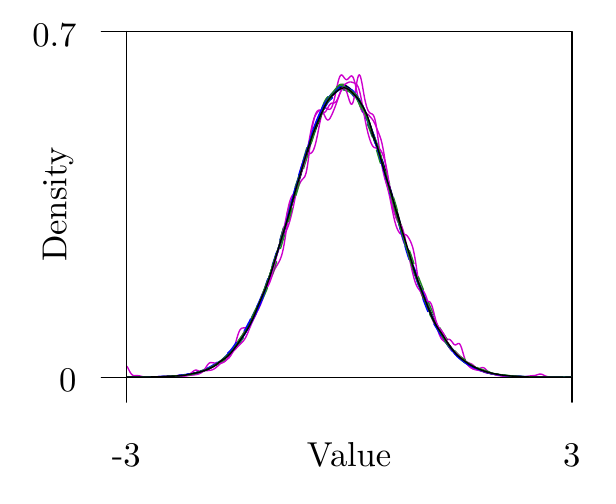}
  \caption{Autocorrelation (left panel) and kernel-density estimate (right panel) of the estimates of the first component of $x_{1}$ for the state-space model in dimension~$d=100$ with $T=10$ observations. Obtained from three independent runs of each of the various Gibbs samplers comprising $500\,000$ iterations (of which the initial $10$~\% were discarded as burn-in) and using $N=100$ particles. Here, $\theta$ was fixed to the true parameters throughout each run. The autocorrelations shown on the r.h.s.\ are averages over the three independent runs of each algorithm. \textit{Note:} the conditional (bootstrap) \gls{PF} almost never managed to update the states: the corresponding kernel-density estimates were therefore not meaningful and are hence suppressed.}
  \label{fig:gibbs_acf_state_components_dimX_100}
 \end{figure}

\begin{figure}[H]
 \centering
 \includegraphics[trim=1cm 0cm 0cm 0.25cm]{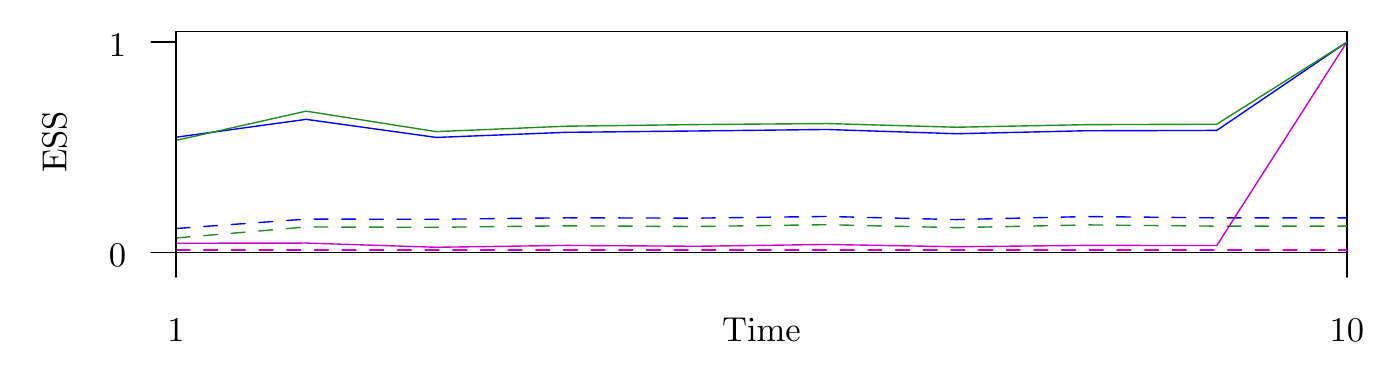}
  \caption{Average \gls{ESS} for the same setting and colour-coding as in Figure~\ref{fig:gibbs_acf_state_components_dimX_100}. The results are averaged over three independent runs of each algorithm. It is worth noting that the \gls{ESS} does not take the autocorrelation of the state-estimates (over iterations of the (particle) \gls{PMCMC} chain) into account and so may flatter \glspl{MCMCPF} to an extent but does illustrate the lessening of weight degeneracy within the particle set which they achieve.}
  \label{fig:gibbs_ess_dimX_100}
  \vspace{3ex}
 \includegraphics[trim=1cm 0cm 0cm 0.25cm]{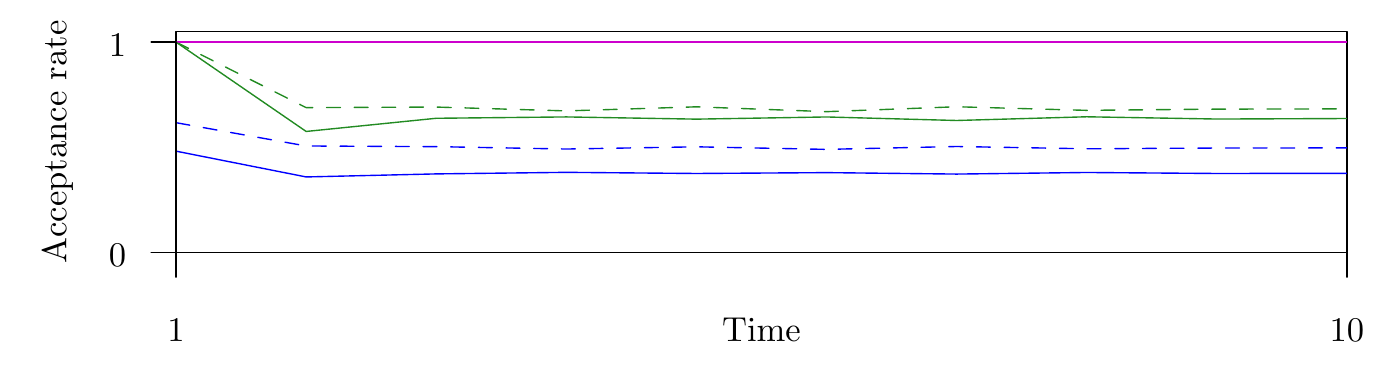}
  \caption{Average acceptance rates for the \gls{MH} kernels $\overline{R}_{\theta, t}(\,\cdot\,|\,\cdot\,;\mathbf{x}_{t-1})$ and $R_{\theta, t}(\,\cdot\,|\,\cdot\,;\mathbf{x}_{t-1})$ for same setting and colour-coding as in Figure~\ref{fig:gibbs_acf_state_components_dimX_100}. Note that standard \glspl{PF} can always be interpreted as using a \gls{MH} kernel that proposes \gls{IID} samples from its invariant distribution so that the acceptance rate is always $1$ in this case. Again the acceptance rates are averaged over three independent runs of each algorithm.}
  \label{fig:gibbs_acceptance_rates_dimX_100}
\end{figure}

Figure~\ref{fig:gibbs_acf_state_components_dimX_100} shows the autocorrelation of estimates of the first component of $x_1$ obtained from various \gls{PG} samplers for model dimension $d=100$. For the moment, we have kept $\theta$ fixed to the true values. It appears that in high dimensions, the conditional \glspl{PF} with \gls{MCMC} moves are able to outperform standard conditional \glspl{PF}. Note that although, unsurprisingly, the best performance is obtained with the \gls{MCMCFAAPF}, the simpler \gls{MCMCPF} is able to substantially outperform the approach based upon a standard \gls{PF}. This is supported by Figure~\ref{fig:gibbs_ess_dimX_100} which shows that the conditional \glspl{PF} with \gls{MCMC} moves lead to a higher estimated \gls{ESS} in this setting. The acceptance rates associated with the \gls{MH} kernels are shown in Figure~\ref{fig:gibbs_acceptance_rates_dimX_100}. 

\begin{figure}[H]
  \noindent{}
  \centering
  \begin{subfigure}[t]{\linewidth}
    \centering
     \includegraphics[trim = 1cm 0cm 0cm 0.25cm]{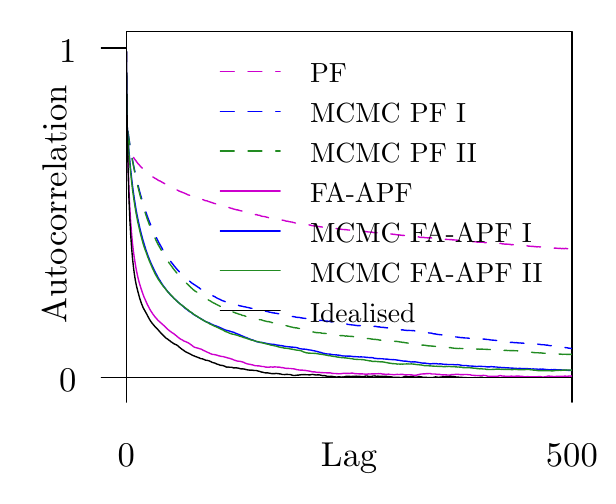}
     \includegraphics[trim = 0cm 0cm 0cm 0.25cm]{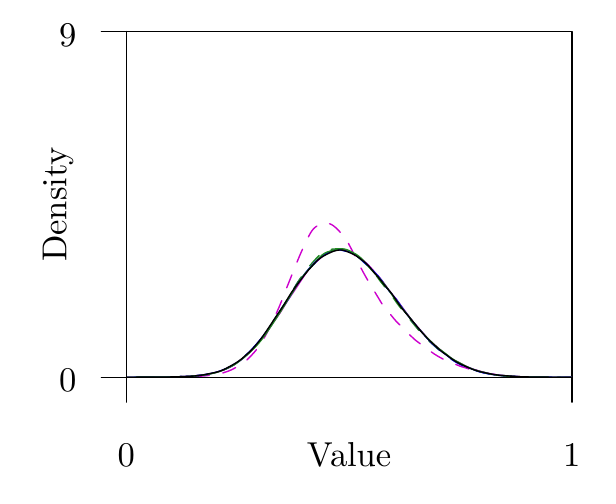}
     \caption{Model dimension~$d=25$.}
  \end{subfigure}
  \begin{subfigure}[t]{\linewidth}
    \centering
     \includegraphics[trim = 1cm 0cm 0cm 0.25cm]{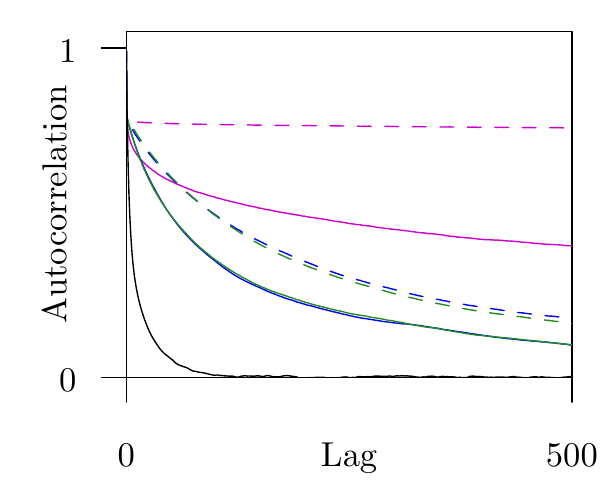}
     \includegraphics[trim = 0cm 0cm 0cm 0.25cm]{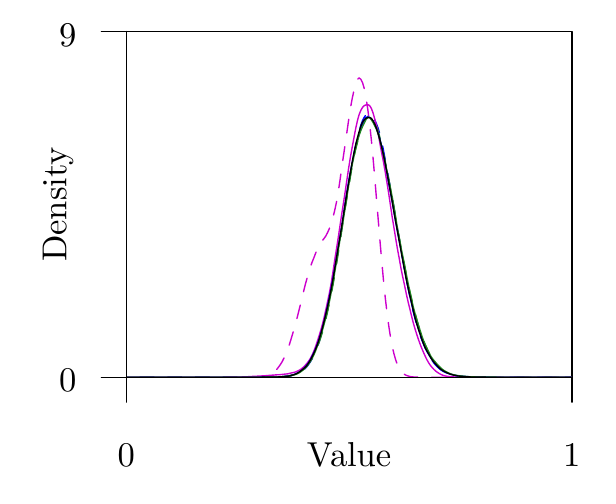}
     \caption{Model dimension~$d=100$.}
  \end{subfigure}
  \caption{Autocorrelation (left panel) and kernel-density estimate (right panel) of the estimates of Parameter~$a_0$ for $T=10$ observations. Obtained one run of the various Gibbs samplers comprising $10^6$ iterations (of which the initial $10$~\% were discarded as burn-in) and using $N=100$ particles. The autocorrelations shown on the l.h.s.\ are averages over the two independent runs of each algorithm.}
  \label{fig:gibbs_acf_parameters}
\end{figure}

We conclude this section by showing (in Figure~\ref{fig:gibbs_acf_parameters}) simulation results for the estimates of Parameter~$a_0$ obtained from the various \gls{PG} samplers. The  \gls{MH} kernel which updates $\theta$ was employed $100$ times per iteration, i.e. $100$ times between each conditional \gls{PF} update of the latent states as the former is relatively computationally cheap compared to the latter. 

Note that as indicated by the kernel-density estimates in the right panel of Figure~\ref{fig:gibbs_acf_parameters}, the Gibbs sampler based around the \gls{PF} did not manage to sufficiently explore the support of the posterior distribution within the number of iterations that we fixed. This lack of convergence also caused the comparatively low empirical autocorrelation of the \gls{PG} chains based around the (bootstrap) \gls{PF} in the left panel of Figure~\ref{fig:gibbs_acf_parameters}: as the chain did not sufficiently traverse support of the target distribution -- due to poor mixing of the state-updates as illustrated in Figure~\ref{fig:gibbs_acf_state_components_dimX_100} -- the empirical autocorrelation shown in Figure~\ref{fig:gibbs_acf_parameters} is a poor estimate of the (theoretical) autocorrelation of the chain. More specifically, the former greatly underestimates the latter. 

The estimates of the other parameters behaved similarly and the results for $(a_1, \sigma, \tau)$  are therefore omitted.

%
%
%
%

\section{Discussion}
\glsresetall

In this work, we have discussed the connections between the \gls{PMCMC} and \gls{EHMM} methodologies and have obtained novel Bayesian inference algorithms for state and parameter estimation in state-space models. We have compared the empirical performance of the various \gls{PMCMC} and \gls{EHMM} algorithms on a simple high-dimensional state-space model. We have found that a properly tuned conditional \gls{PF} which employs local \glsdesc{MH} moves proposed in \citet{ShestopaloffNeal2016} can dramatically outperform the standard conditional \glspl{PF} in high dimensions. Additionally, by formally establishing that \gls{PMCMC} and the (alternative) \gls{EHMM} methods can be viewed as a special case of a general \gls{PMCMC} framework, we have derived both \glsdesc{BS} and \glsdesc{AS} for this general framework. This provides a promising strategy for extending the range of applicability of \glsdesc{PG} algorithms as well as providing a novel class of \glspl{PF} which might be useful.

There are numerous other potential extensions of these ideas. For instance, many existing extensions of standard \gls{PMCMC} methods could also be considered for the alternative \gls{EHMM} methods, e.g.\ incorporating gradient-information into the parameter proposals $q(\theta, {\theta'})$ or exploiting correlated pseudo-marginal ideas \citep{Deligiannidis2015}. Clearly, further generalisation of the target distribution and associated algorithms introduced here are possible. Many other processes for simulating from an extended target admitting a single random trajectory with the correct marginal distribution are possible, e.g.\ along the lines of \citet{LindstenJohansenNaesseth2014}.


\section*{Acknowledgements}
Arnaud Doucet's research is partially supported by the \gls{EPSRC}, grants EP/K000276/1, EP/K009850/1 and by the Air Force Office of Scientific Research/Asian Office of Aerospace Research and Development, grant AFOSRA/AOARD-144042. Axel Finke was partially supported by the \gls{EPSRC} under grants EP/I017984/1 and EP/K020153/1.

\setlength{\bibsep}{3pt plus 0.3ex}
\bibliography{ehmm}

\appendix

\section{Special cases of the general PMCMC algorithm}
\label{subsec:special_cases}
\glsunset{MCMC}
\glsunset{PF}
\glsunset{FAAPF}
\glsunset{APF}
\glsunset{MCMCPF}
\glsunset{MCMCFAAPF}
\glsunset{MCMCAPF}

In this appendix, we show that all \gls{PMCMC} and alternative \gls{EHMM} methods described this work can be recovered as special cases of the general \gls{PMCMC} framework from Section~\ref{sec:general_pmcmc}. For completeness, we explicitly derive all algorithms as special cases of the general framework even though \gls{PMCMC} methods based around the (bootstrap) \gls{PF} and \gls{FAAPF} were already shown to be special cases of \gls{PMCMC} methods based around the general \gls{APF} and even though, alternative \gls{EHMM} methods based around the \gls{MCMCPF} and \gls{MCMCFAAPF} were already shown to be special cases of alternative \gls{EHMM} methods based around the \gls{MCMCAPF}.

\begin{description}[leftmargin=0pt, labelindent=0pt]
 \item[(Bootstrap) \gls{PF}.]
  In this case, $\psi_{\theta,1}(\mathbf{z}_1) = \prod_{i=1}^N \mu_\theta(x_1^i) = \prod_{i=1}^N \bar{\rho}_{\theta, 1}(x_1^i)$, and, for $t>1$, 
  \begin{equation}
   \psi_{\theta,t}(\mathbf{z}_t|\mathbf{z}_{1:t-1}) = \prod_{i=1}^N \frac{g_\theta(y_{t-1}|x_{t-1}^{a_{t-1}^i})}{\sum_{j=1}^N g_\theta(y_{t-1}|x_{t-1}^j)} f_\theta(x_t^i|x_{t-1}^{a_{t-1}^i}) = \prod_{i=1}^N \bar{\rho}_{\theta, t}(x_t^i, a_{t-1}^i|\mathbf{x}_{t-1}),                                                                                                        
  \end{equation}
  while $\gamma_{\theta,t}(x_{1:t}) \coloneqq p_\theta(x_{1:t}, y_{1:t})$, for any $t \leq T$. This implies that $\tilde{v}_{\theta,t}^{i} = \frac{1}{N} g_\theta(y_t|x_t^{i}) \prod_{n=1}^{t-1} \frac{1}{N} \sum_{j=1}^N g_\theta(y_n|x_n^j)$, so that we obtain $q_\theta(i|\mathbf{z}_{1:T}) = g_\theta(y_T|x_T^i) / \sum_{j=1}^N g_\theta(y_T|x_T^j)$ and $\hat{p}_\theta(y_{1:T}) = \prod_{t=1}^{T} \frac{1}{N} \sum_{i=1}^N g_\theta(y_t|x_t^i)$, as stated in Section~\ref{sec:pmcmc}.

\item [\gls{FAAPF}.] In this case, $\psi_{\theta,1}(\mathbf{z}_1) = \prod_{i=1}^N p_\theta(x_1^i|y_1) = \prod_{i=1}^N \rho_{\theta,1}(x_1^i)$, and, for $t>1$, 
  \begin{equation}
   \psi_{\theta,t}(\mathbf{z}_t|\mathbf{z}_{1:t-1}) = \prod_{i=1}^N \frac{g_\theta(y_{t}|x_{t-1}^{a_{t-1}^i})}{\sum_{j=1}^N g_\theta(y_{t}|x_{t-1}^j)} p_\theta(x_t^i|x_{t-1}^{a_{t-1}^i}, y_t) = \prod_{i=1}^N \rho_{\theta,t}(x_t^i, a_{t-1}^i|\mathbf{x}_{t-1}),                                                                                                    
  \end{equation}
    while $\gamma_{\theta,t}(x_{1:t}) \coloneqq p_\theta(x_{1:t}, y_{1:t}) p_\theta(y_{t+1}|x_t)$, for $t < T$, and $\gamma_{\theta,T}(x_{1:T}) \coloneqq p_\theta(x_{1:T}, y_{1:T})$.  This implies that $\tilde{v}_{\theta,t}^{i} = \frac{1}{N} p_\theta(y_1) \prod_{n=2}^t \frac{1}{N} \sum_{j=1}^N p_\theta(y_n|x_{n-1}^j)$, so that we obtain the selection probability $q_\theta(i|\mathbf{z}_{1:T}) = 1/N$ and the marginal-likelihood estimate $\hat{p}_\theta(y_{1:T}) = p_\theta(y_1) \prod_{t=2}^T \frac{1}{N} \sum_{i=1}^N p_\theta(y_t|x_{t-1}^i)$, as stated in Section~\ref{sec:pmcmc}.

\item[General \gls{APF}.] In this case, $\psi_{\theta,1}(\mathbf{z}_1) = \prod_{i=1}^N q_{\theta,1}(x_1^i) = \prod_{i=1}^N \rho_{\theta,1}^{\mathbf{q}_\theta}(x_1^i)$, and, for $t>1$, 
  \begin{equation}
   \psi_{\theta,t}(\mathbf{z}_t|\mathbf{z}_{1:t-1}) = \prod_{i=1}^N \frac{v_{\theta, t-1}^{a_{t-1}^i}}{\sum_{j=1}^N v_{\theta, t-1}^j} q_{\theta,t}(x_t^i|x_{t-1}^{a_{t-1}^i}) = \prod_{i=1}^N \rho_{\theta,t}^{\mathbf{q}_\theta}(x_t^i, a_{t-1}^i|\mathbf{x}_{t-2:t-1},  \mathbf{a}_{t-2}),
  \end{equation}
    while $\gamma_{\theta,t}(x_{1:t}) \coloneqq p_\theta(x_{1:t}, y_{1:t}) \tilde{p}_\theta(y_{t+1}|x_t)$, for $t < T$, and $\gamma_{\theta,T}(x_{1:T}) \coloneqq p_\theta(x_{1:T}, y_{1:T})$.  This implies that $\tilde{v}_{\theta,t}^{i} = \frac{1}{N} v_{\theta,t}^i \prod_{n=1}^{t-1} \frac{1}{N} \sum_{j=1}^N v_{\theta,n}^j$, so that we obtain the selection probability $q_\theta(i|\mathbf{z}_{1:T}) = v_{T,\theta}^i / \sum_{j=1}^N v_{T,\theta}^j$ and the marginal-likelihood estimate  $\hat{p}_\theta(y_{1:T}) = \prod_{t=1}^T \frac{1}{N} \sum_{i=1}^N v_{t,\theta}^i$, as stated in Section~\ref{sec:pmcmc}.
    
\item[\gls{MCMCPF}.]
 In this case, $\psi_{\theta,1}(\mathbf{z}_1) = \bar{\rho}_{\theta,1}(x_1^1) \prod_{i=2}^N \overline{R}_{\theta,1}(x_1^i|x_1^{i-1})$, and, for $t>1$, 
  \begin{equation}
   \psi_{\theta,t}(\mathbf{z}_t|\mathbf{z}_{1:t-1}) = \bar{\rho}_{\theta,t}(x_t^1, a_{t-1}^1|\mathbf{x}_{t-1}) \prod_{i=2}^N \overline{R}_{\theta,t}(x_t^i, a_{t-1}^i|x_t^{i-1}, a_{t-1}^{i-1}; \mathbf{x}_{t-1}),                                                                                                        
  \end{equation}
  while $\gamma_{\theta,t}(x_{1:t})$, $q_\theta(b_T|\mathbf{z}_{1:T})$ and $\hat{p}_\theta(y_{1:T})$ are the same as for \gls{PMCMC} methods using the bootstrap \gls{PF}.

  \item[\gls{MCMCFAAPF}.] In this case, $\psi_{\theta,1}(\mathbf{z}_1) = \rho_{\theta,1}(x_1^1) \prod_{i=2}^N R_{\theta,1}(x_1^i|x_1^{i-1})$, and, for $t>1$, 
  \begin{equation}
   \psi_{\theta,t}(\mathbf{z}_t|\mathbf{z}_{1:t-1}) = \rho_{\theta,t}(x_t^1, a_{t-1}^1|\mathbf{x}_{t-1}) \prod_{i=2}^N R_{\theta,t}(x_t^i, a_{t-1}^i|x_t^{i-1}, a_{t-1}^{i-1}; \mathbf{x}_{t-1}),                                                                                                        
  \end{equation}
  while $\gamma_{\theta,t}(x_{1:t})$, $q_\theta(b_T|\mathbf{z}_{1:T})$ and $\hat{p}_\theta(y_{1:T})$ are the same as for \gls{PMCMC} methods using the \gls{FAAPF}.

  \item[\gls{MCMCAPF}.] In this case, $\psi_{\theta,1}(\mathbf{z}_1) = \rho_{\theta,1}^{\mathbf{q}_\theta}(x_1^1) \prod_{i=2}^N R_{\theta,1}^{\mathbf{q}_\theta}(x_1^i|x_1^{i-1})$, and, for $t>1$, 
  \begin{equation}
   \psi_{\theta,t}(\mathbf{z}_t|\mathbf{z}_{1:t-1}) = \rho_{\theta,t}^{\mathbf{q}_\theta}(x_t^1, a_{t-1}^1|\mathbf{x}_{t-2:t-1},  \mathbf{a}_{t-2}) \prod_{i=2}^N R_{\theta,t}^{\mathbf{q}_\theta}(x_t^i, a_{t-1}^i|x_t^{i-1}, a_{t-1}^{i-1}; \mathbf{x}_{t-2:t-1},  \mathbf{a}_{t-2}),
  \end{equation}
   while $\gamma_{\theta,t}(x_{1:t})$, $q_\theta(b_T|\mathbf{z}_{1:T})$ and $\hat{p}_\theta(y_{1:T})$ are the same as for \gls{PMCMC} methods using the general APF.
\end{description}
\end{document}